\documentclass{pasj00}
\draft

\begin{document}
\SetRunningHead{W.-S. Jeong et al.}{ASTRO-F/FIS Observing
Simulation}
\Received{2002 July 15}
\Accepted{2003 March 7}

\title{ASTRO-F/FIS Observing Simulation: \\ Detection Limits for Point Sources}

\author{
    Woong-Seob \textsc{Jeong}\altaffilmark{1},
    Soojong \textsc{Pak}\altaffilmark{1,2,3},
    Hyung Mok \textsc{Lee}\altaffilmark{1},
    Takao \textsc{Nakagawa}\altaffilmark{3},
    Jungjoo \textsc{Sohn}\altaffilmark{1}, \\
    Insun \textsc{Ahn}\altaffilmark{1},
    Issei \textsc{Yamamura}\altaffilmark{3},
    Masaru \textsc{Watanabe}\altaffilmark{3},
    Mitsunobu \textsc{Kawada}\altaffilmark{4} and
    Hiroshi \textsc{Shibai}\altaffilmark{4}
}

\altaffiltext{1}{Astronomy Program in Graduate School of Earth and Enviromental
Sciences, \\ Seoul National University, Shillim-Dong, Kwanak-Gu, Seoul 151-742,
South Korea} \email{jeongws@astro.snu.ac.kr, soojong@astro.snu.ac.kr,
hmlee@astro.snu.ac.kr, \\jjsohn@astro.snu.ac.kr, ais@astro.snu.ac.kr}
\altaffiltext{2}{Korea Astronomy Observatory, Whaam-Dong, Youseong-Gu, Taejeon
305-348, South Korea} \altaffiltext{3}{Institute of Space and Astronautical
Science, Yoshinodai 3-1-1, Sagamihara, Kanagawa 229-8510, Japan}
\email{nakagawa@ir.isas.ac.jp, yamamura@ir.isas.ac.jp, watanabe@ir.isas.ac.jp}
\altaffiltext{4}{Graduate School of Science, Nagoya University, Furo-cho,
Chigusa-ku, Nagoya 464-8602, Japan} \email{kawada@u.phys.nagoya-u.ac.jp,
shibai@phys.nagoya-u.ac.jp}

%


\maketitle

\begin{abstract}
We describe the observing simulation software FISVI (FIS Virtual Instrument),
which was developed for the Far-Infrared Surveyor (FIS) that will be on the
Japanese infrared astronomy mission ASTRO-F. The FISVI has two purposes: one is
to check the specifications and performances of the ASTRO-F/FIS as a whole; the
other is to prepare input data sets for the data analysis softwares prior to
launch. In the FISVI, special care was taken by introducing the ``Compiled PSF
(Point Spread Function)'' to optimise inevitable, but time-consuming,
convolution processes. With the Compiled PSF, we reduce the computation time by
an order of magnitude. The photon and readout noises are included in the
simulations. We estimate the detection limits for point sources from the
simulation of virtual patches of the sky mostly consisting of distant galaxies.
We studied the importance of source confusion for simple power-law models for
$N(>S)$, the number of sources brighter than $S$. We found that source
confusion plays a dominant role in the detection limits only for models with
rapid luminosity evolution for the galaxy counts, the evolution of which is
suggested by recent observations.
\end{abstract}

\section{Introduction}\label{sec:introduction}

The FIS (Far-Infrared Surveyor) is one of the focal plane instruments of the
ASTRO-F mission (previously known as IRIS) (\cite{mura98}; \cite{shib00};
\cite{naka01}). The ASTRO-F satellite will be launched into a sun-synchronous
orbit at an altitude of 750~km, which corresponds to an orbital period of 100
min. The telescope, which is cooled down to 5.1--5.8 K, has a 67 cm primary
mirror. The major task of this mission is to carry out an all-sky survey across
the 50--200 $\mu$m range. The basic parameters of the ASTRO-F/FIS are
summarized in table \ref{tab_spec_fis} (see also Kawada 2000).

    \begin{table}
    \begin{center}
    \caption{Specifications of the FIS.}
       \vskip 3 truemm
    \begin{tabular}{l c c c c c} \hline\hline\noalign{\smallskip}
     & Wavelength range & Array size & Pixel size & Pitch size & Sampling rate \\
     Band & ($\mu$m) & (pixel) & (arcsec) & (arcsec) & (Hz)\\
    \hline\noalign{\smallskip}
     WIDE-L& 110 $-$ 200 & 15 $\times$ 3 & 44.2 & 49.1 & 15.2 \\
     N170 & 150 $-$ 200 & 15 $\times$ 2 & 44.2 & 49.1 & 15.2 \\
     WIDE-S & 50 $-$ 110 & 20 $\times$ 3 & 26.8 & 29.5 & 22.8 \\
     N60 & 50 $-$ 75 & 20 $\times$ 2 & 26.8 & 29.5 & 22.8 \\
    \hline
    \end{tabular} \label{tab_spec_fis}
    \end{center}
    \end{table}

ASTRO-F/FIS will bring data with much higher sensitivity and angular resolution
than those of IRAS (see Kawada 2000 for detailed comparison). Such data sets
will be of great value for many areas of astrophysics, including cosmology,
galaxy evolution, interstellar medium, and asteroids.

Generally speaking, the hardware characteristics of each component in a space
mission can be measured in the laboratory. However, it is very difficult to
make end-to-end tests of a mission in the laboratory. Hence, based on data
measured for each component, numerical simulations are frequently used to
understand the instrument performances as a whole (e.g., \cite{garc98};
\cite{boggs01}). Moreover, the complicated interplay between the celestial
sources and hardware specifications can be studied only by a simulation prior
to the launch.

We have constructed a software simulator called the FISVI representing Virtual
Instrument of the FIS, that can simulate the data stream of ASTRO-F/FIS
(\cite{jeong00}). This work is an extension of initial work by \citet{matsu01}.
The purposes of the FISVI are : (1) to confirm the performance of the hardware
as a whole and (2) to generate simulated FIS survey data sets as inputs for
data-reduction software prior to launch.

One of the key questions regarding the performance of ASTRO-F is the effective
detection limit for faint sources. Depending on the size of the sources
compared to the beam size of ASTRO-F, the source can be either extended or
point-like, and the detection limits depend on the nature of the sources. In
the present work, we only consider point sources.

There are several factors contributing to the detection limits. The sensitivity
of the detectors and the entire telescope system allows only sources brighter
than a certain threshold to be reliably measured. Since the photons follow
Poisson statistics, the background photons due to the sky brightness as well as
the telescope emission should fluctuate, and a meaningful detection of a source
can be made only if the signal from the source exceeds the level of the
fluctuations. The sky confusion noise by the cirrus emission causes an
uncertainty in the determination of the source flux, due to the variation of
the sky brightness (\cite{herb98}; \cite{kiss01}). The readout process also
adds more fluctuations. Moreover, the measurement of the brightness of a source
can be further influenced by neighboring sources if more than one source lies
within a single beam of the telescope. The final detection limit should thus
depend on the performance of the entire system, the brightness of sky and
telescope emission, readout process, and the distribution of sources as a
function of the flux.

There have been a number of estimates of detection limits based on the
available laboratory data (e.g., \cite{kawada98}, 2000) using simple
calculations. Clearly, a more realistic estimation can be made by using
numerical simulations. In the present work, we carried out simulations of the
ASTRO-F/FIS observations under several different circumstances in order to
obtain still more reliable detection limits which can be used to design
scientific projects.

The present paper is organised as follows. In section \ref{sec:struc_fisvi}, we
briefly describe the design and the structure of the FISVI. In section
\ref{sec:imgrc_pa}, we explain how we obtain the observed images based on the
simulated data set. In section \ref{sec:calc_detlim}, we make estimate on the
detection limits of the ASTRO-F/FIS under various circumstances. First, we
estimate the detection limits of a single isolated point source while
considering only photon and readout noises. Also, we estimate the confusion
noise (\cite{cond74}; \cite{fran89}) for distributed sources using a simple
formula. By carrying out aperture photometry to the simulated images, we
finally obtain combined detection limits that include photon, readout, and
confusion noises. The final section summarises our conclusions.

\section{Structure of FISVI Software}\label{sec:struc_fisvi}

\begin{figure}
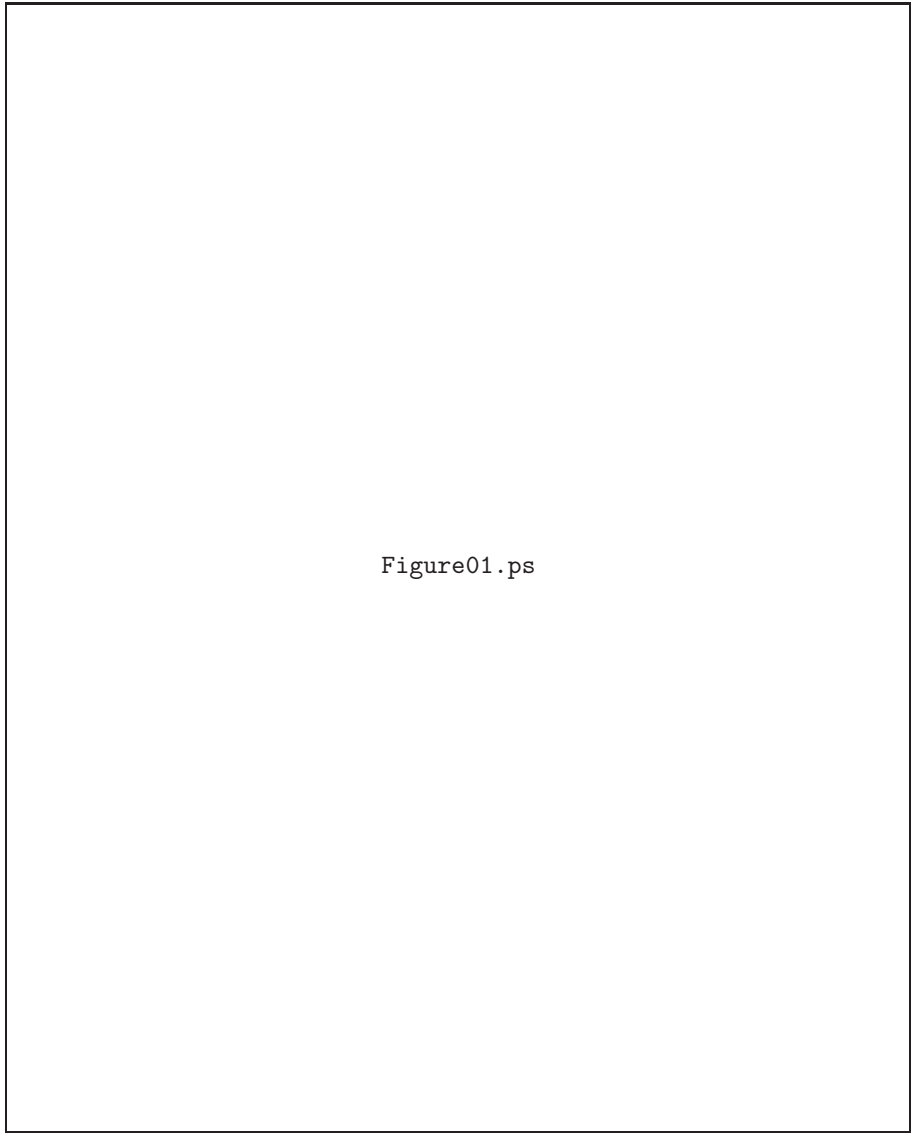

  \begin{center}
    \FigureFile(120mm, 150mm){Figure01.ps}
  \end{center}
      \caption{Flow charts of FISVI. The left chart shows
      a straightforward procedure based on the realistic photon path,
      where repeated calculations would be necessary over the wavelength
      grids (``$\lambda$ Convolution''). The right chart, on the
      other hand, shows the accelerated algorithm using the Compiled PSF
      for the FISVI.}
   \label{fig_flow}
\end{figure}

The algorithm of the FISVI software is shown in figure \ref{fig_flow}. The
input data file provides the coordinates and fluxes of the sources in the sky.
Although the sources would appear either point-like or extended, we concentrate
on point sources in this paper. The software first makes images on the focal
plane by convolving the point sources and the Point Spread Function (PSF) of
the telescope and the instrument. The software generates time-series data for
each pixel by simulating the scanning procedure of the ASTRO-F/FIS survey mode
observations.

Since the PSF, the filter transmission, and the detector response depend on the
wavelength of incoming photons, we need to do repeated calculations (procedures
boxed in the left panel of figure \ref{fig_flow}) for different wavelengths
within the individual FIS bands, as shown in the left panel of figure
\ref{fig_flow}. To elude this and speed up the procedure, we introduce the
Compiled PSF in this work, with which we can perform this scanning procedure at
once, as shown in the simplified flow chart in the right panel of figure
\ref{fig_flow}. A more detailed discussion on the gains in the computational
time and possible errors due to the use of the Compiled PSF are presented in
appendix \ref{sec:comp_PSF}.

The readout values for each pixel are represented by a series of integrated
charges taken over the area covered by the pixel, sampled at regular time
intervals.  The integrated charges are set to zero at every reset interval. The
time series of the integrated charges are differentiated to obtain the charges
accumulated during the sampling interval (see appendices \ref{sec:pos} and
\ref{sec:sampling} for detailed process). We also generate the photon and
readout noise and include them to be part of readout values. A more detailed
discussion on the implementation of noise is presented in subsection
\ref{sec:detlim_single}. The time-series data are converted into the brightness
distribution on the sky, and are used to reconstruct the images, as described
in section \ref{sec:imgrc_pa}.

\section{Image Reconstruction} \label{sec:imgrc_pa}
\begin{figure}
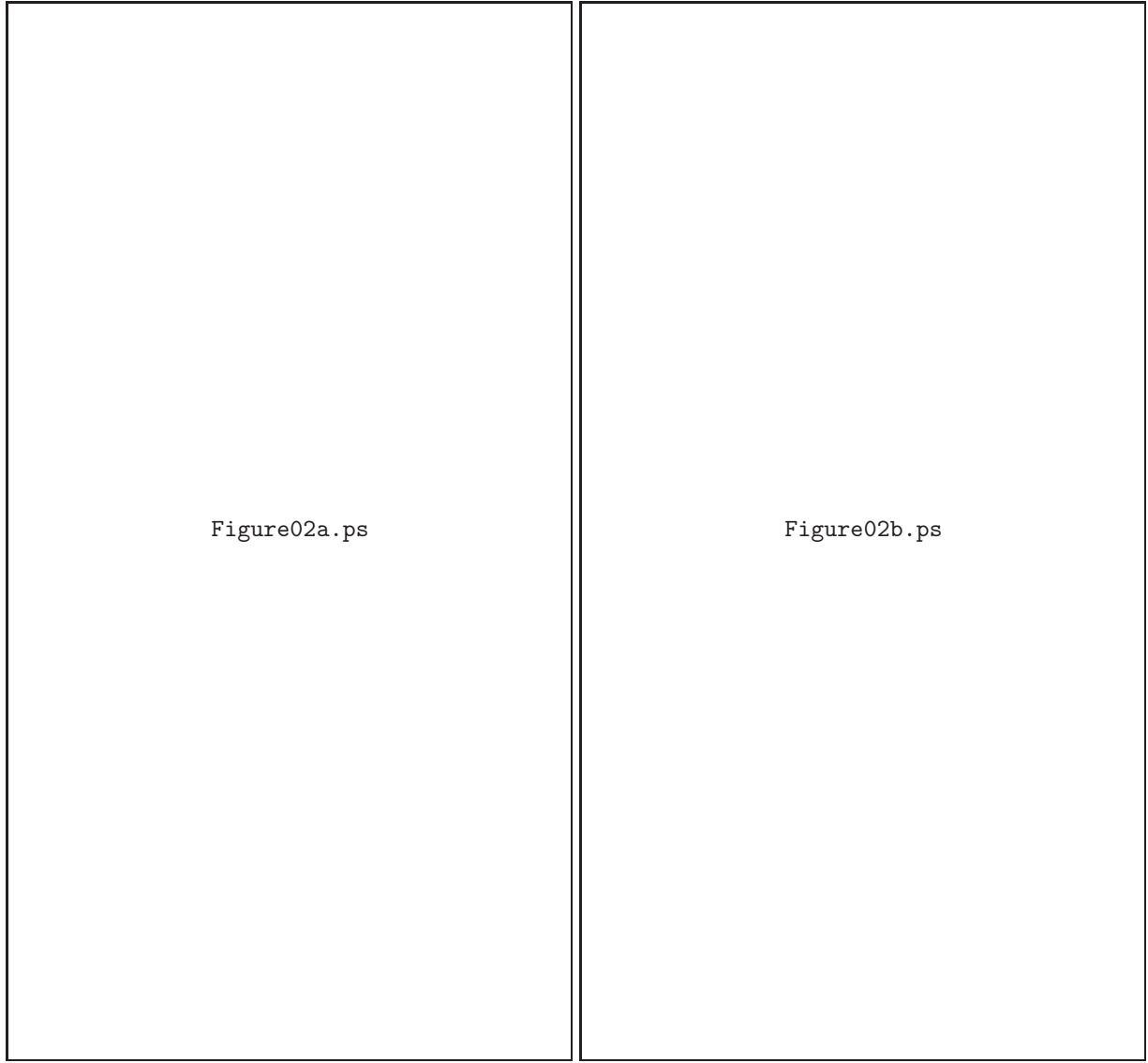

  \begin{center}
    \FigureFile(80mm, 150mm){Figure02a.ps}
    \FigureFile(80mm, 150mm){Figure02b.ps}
  \end{center}
   \caption{Example of a series of readout values, which corresponds to
   the integrated charges since the last reset [see equation (\ref{eq_totcharge})],
   of a WIDE-L pixel that scans through a point source (left panel).
   The differentiation of the integrated charges as shown in the
   right panel corresponds to the signal obtained during a sampling
   interval by one detector pixel passing the image of a point source.
   The sampling interval was $14\farcs2$, corresponding to
   the 15.2 Hz readout (see table \ref{tab_spec_fis}).}
   \label{fig_single_int}
\end{figure}

The FISVI generates time-series data for each pixel. In figure
\ref{fig_single_int}, we show a series of readout values of a pixel that scans
across a point source. No reset was applied during the readout sequence shown
in this figure because the reset time interval is usually much longer than the
passage of a Compiled PSF over a point source. The differentiation (subtraction
of adjacent sampling points) of this curve gives the signal obtained during a
sampling interval by one detector pixel, which is shown in the lower panel of
figure \ref{fig_single_int}.

\begin{figure}
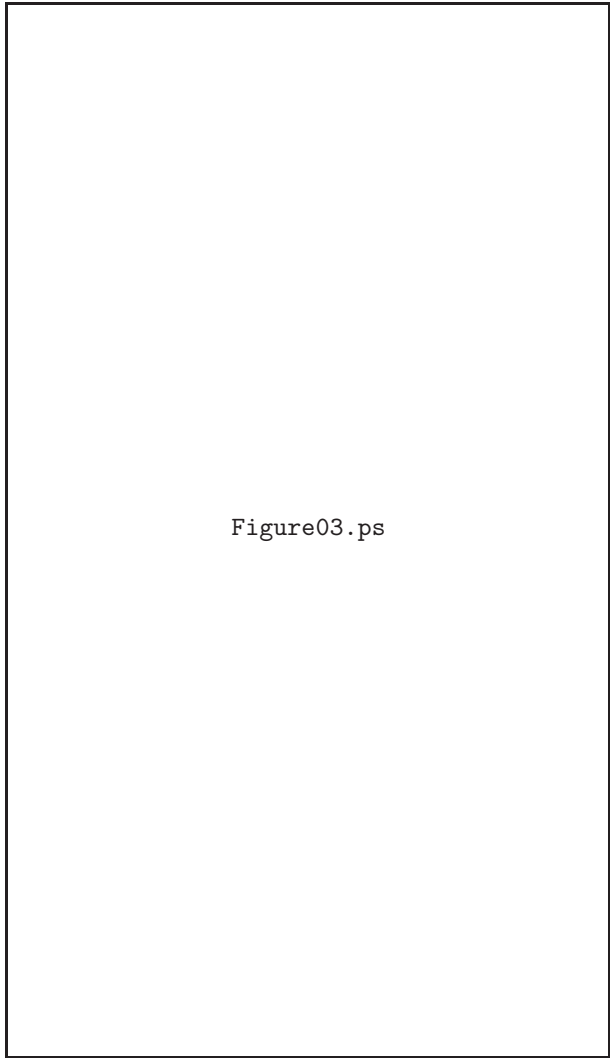

  \begin{center}
    \FigureFile(80mm, 140mm){Figure03.ps}
  \end{center}
    \caption{Schematic figure for image reconstruction by pixel
    averages. At any given point, we take the average of the pixel
    readouts that were covered by those pixels. In the figure,
    the darker area means the area that was covered more.}
    \label{fig_sum}
\end{figure}

The pixel readouts can be used to reconstruct the images. In the
current implementation of the FISVI, following method was used to
generate the image.
In order to reconstruct the image, we assume that pixel value represents the
uniform intensity over the pixel surface. This means that a particular point
can be covered by more than one readout. We always take the average values of
multiple readouts in order to construct images (see figure \ref{fig_sum}). Due
to the convolution of the image with the pixel size, the output image will be
blurred slightly.


\section{Estimations of FIS Performance}\label{sec:calc_detlim}

An estimation of the detection limits for the planned mission is very
important. For ASTRO-F, the detection limits were estimated by using analytic
methods (Kawada 1998, 2000). In the present work, we made a numerical estimate
for a single point source using the latest information for the detectors and
filters, and compared them with the photometric results on the FISVI generated
images that contain a large number of point sources.

\subsection{Detection Limits for a Single Point Source}
\label{sec:detlim_single}

\begin{figure}
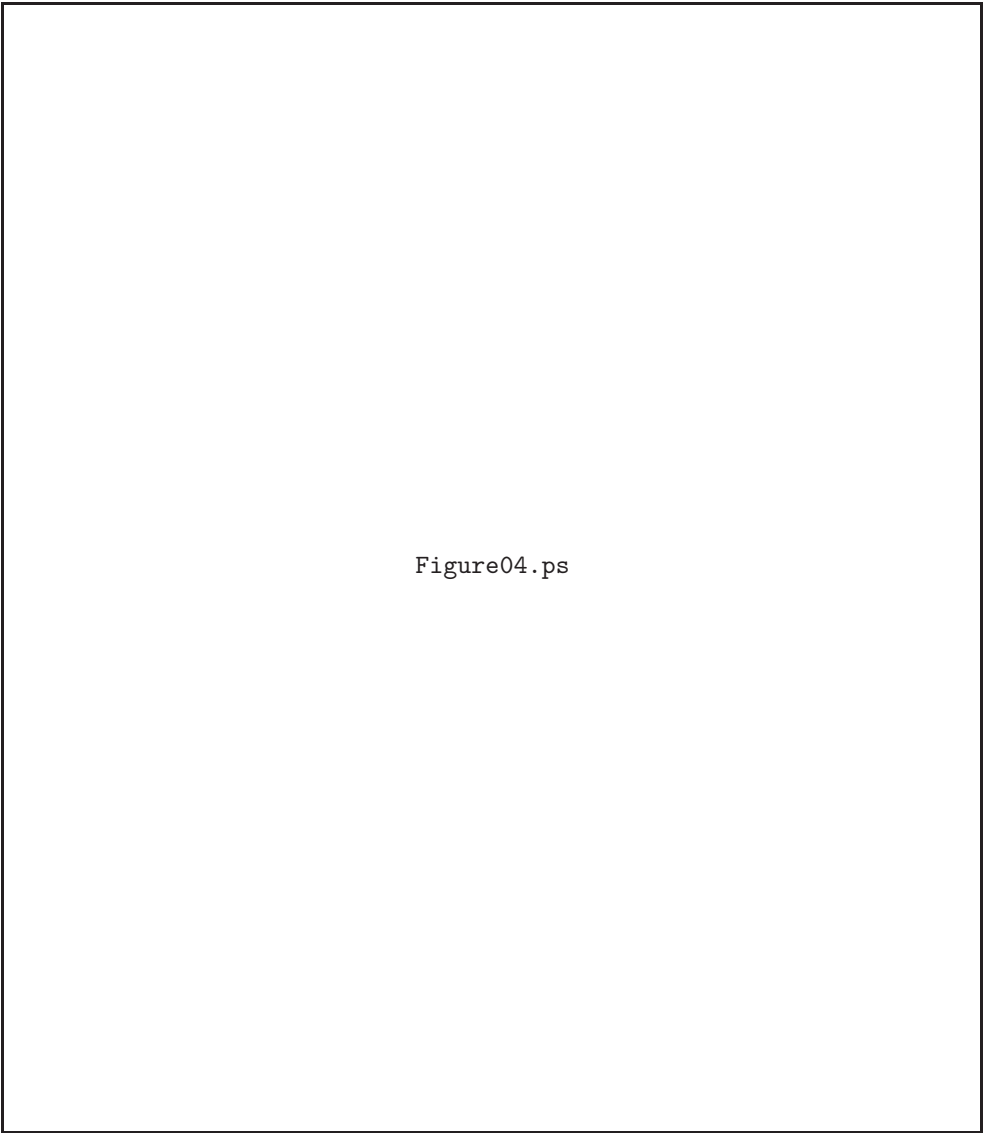

  \begin{center}
    \FigureFile(130mm, 150mm){Figure04.ps}
  \end{center}
   \caption{Assumed background emissions. We consider three components
   for the background emission, i.e., interstellar dust (dotted),
   interplanetary dust (dashed) and telescope emission assuming 6 K
   (dot-dashed line) or 6.5 K black body (long dot-dashed line).
   In our simulations, the telescope temperature is always assumed to be 6 K.}
   \label{bgr_kawada}
\end{figure}

The detection limits for a single point source depend on the level of noise.
There are several sources of noise: photon noise due to the sky background and
thermal emission from the telescope, and readout noise. The sky background
varies significantly from place to place in the sky. On average, the infrared
sky becomes brighter in the Galactic plane, and diminishes toward the Galactic
poles. Within the Galactic plane, the emission from the Galactic center
direction appears to be brighter than towards the anti-center direction.
Because of thermal emission by interplanetary dust particles, the ecliptic
plane is also brighter than the ecliptic pole region. In figure
\ref{bgr_kawada}, we show the assumed surface brightness distribution of
background emissions from the interstellar dust, the interplanetary dust and
the telescope, for the purpose of generating photon noises. These background
emissions from the sky are assumed to correspond to the dark part of the sky
and the sky confusion noise due to the structure of the cirrus emission is not
considered. The telescope temperature is assumed to be 6 K, as a conservative
number. In figure \ref{bgr_kawada}, we also plotted the thermal emission from
the 6.5 K telescope as a comparison. Evidently, the contribution from the
telescope is smaller than that from the interplanetary or interstellar dust as
long as the telescope temperature is lower than 6.5 K for the entire FIS bands.
The sky brightness throughout the spectral region of the FIS varies from 5 to 7
MJy\,sr$^{-1}$. Obviously, we would need to apply a position-dependent
background brightness for more realistic sky simulations, which affects the
photon noise. The incoming photon stream on pixels due to background emission
is assumed to follow Poisson statistics.

The readout circuit also generates uncertainties of the output values, called
readout noise. This type of noise is independent of the sampling rate and the
integration time, and we assumed the total noise in the effective bandwidth at
the first stage of the field effect transistor (FET) gate to be 3 $\mu$V. In
the simulation, we assumed that the readout noise follows Gaussian statistics.

\subsubsection{Simple estimation}\label{sec:analy_est}

The sky brightness throughout the spectral region of the FIS varies from 5 to 7
MJy\,sr$^{-1}$. The integrated photons fluctuate following the Poisson
statistics while the readout process adds readout noise, which is assumed to
follow Gaussian statistics. The r.m.s. fluctuation of voltage across the
integrating charge due to readout noise can be converted to the fluctuation in
the number of charges by

\begin{equation}
D_{\rm rms} = {C~V_{\rm rms} \over e},
\end{equation}
where $C$ is the capacitance of the charge integrators [7 pF for SW (short
wavelength) and 10 pF for LW (long wavelength) bands, respectively], and $e$ is
the elementary charge. The total noise is a combination of photon and readout
noise.

If we assume that a single pixel detector receives the entire photon flux of
the point source, we can obtain the accumulated charge during `the effective
integration time' that elapses until the detector pixel passes through one
point. For a photoconductor, the noise by this photon flux arises from the
sequence of generations and recombinations of photoelectrons. We calculated
this generation-recombination noise (G-R noise), $I_{\rm G-R}$ (\cite{rieke94})
using
\begin{equation}
\langle I_{\rm G-R}^2\rangle = 4e^2 \varphi \eta G^2 df,
\label{eq_gr_noise}
\end{equation}
where $\varphi$ is the photon flux, $\eta$ is the quantum efficiency, $G$ is
the photoconductive gain, and $df$ is the effective bandwidth. We assumed that
the source has the SED of a 40 K blackbody. The 5$\sigma$ detection limits
computed in this way for all FIS bands are shown in table \ref{tab_analy_est}.
Also shown in this table is the relative importance of the photon and readout
noise. In all cases, the readout noise is more important than the photon noise,
with narrow bands (N170 and N60) being more dominated by readout noise.

    \begin{table}
    \begin{center}
    \caption{Simple estimates of 5$\sigma$ detection of single pixels and
    the ratios of photon-to-readout noises.}
    \vskip 3 truemm
    \begin{tabular}{l c c} \hline\hline\noalign{\smallskip}
     & 5$\sigma$ Det. Limit (mJy)$^\mathrm{*}$
     & $\sigma_{\rm r}/\sigma_{\rm ph}^\mathrm{\dagger}$ \\
    \hline\noalign{\smallskip}
     WIDE-L& 39 & 1.3 \\
     N170 & 76 & 1.8 \\
     WIDE-S & 20 & 1.6 \\
     N60 & 52 & 2.5 \\
    \hline
    \end{tabular} \label{tab_analy_est}
    \begin{list}{}{}
    \item[$^{\mathrm{*}}$] Average flux density in the bandwidth.
    \item[$^{\mathrm{\dagger}}$] Readout-to-photon noise ratio.
    \end{list}
    \end{center}
    \end{table}

\subsubsection{Estimation using scanning simulations of a single pixel}\label{sec:scan_est}
We also estimate the detection limits from the detector scanning routines in
the FISVI for a single pixel. The behavior of the readout values as a function
of the sampling sequence is shown in figure \ref{fig_single_int}. The
contribution due to background can be obtained by subtracting the contribution
from the source alone. The expected amount of the fluctuation is proportional
to $G\sqrt{\varphi \eta}$ for a given span of the scanning period of $t_1$ to
$t_2$ [see equation (\ref {eq_gr_noise})]. The total amount of fluctuation of
the readout value due to noises during the same scanning span, $\sigma_{\rm
tot}$, is
\begin{equation}
\sigma_{\rm tot} = \sqrt{\sigma_{\rm ph}^2 + \sigma_{\rm r}^2},
\end{equation}
where $\sigma_{\rm ph}$ and $\sigma_{\rm r}$ are the fluctuation due to the
photon and readout noise, respectively. Here, we assume that the readout noise
is always a constant while the amount of charge fluctuation due to the photon
noise increases as $G\sqrt{\varphi \eta}$, as dictated by the Poisson nature.
For a given brightness of a source, we can obtain the $S/N$ ratio if we specify
$t_1$ and $t_2$. Since the signal (photocurrent) and the photon noise are
proportional to $G$, $S/N$ depends on $\sqrt{\eta}$ on the condition that the
photon noise is the dominant case. From equation (\ref{eq_ms_resp}) and the
assumption $G = 0.9$, we can obtain the quantum efficiency, $\eta$, as 0.17 for
SW and 0.27 for LW detectors, respectively. The determination of $t_2$ and
$t_1$ was done to maximize the $S/N$. We find that this can be done when we
start the scanning at a distance of 2$W_{\rm H}$ and continue until the same
distance in the opposite side, where $W_{\rm H}$ is the full width at half
maximum of the beam patterns (see subsubsection \ref{sec:sim_conf} for
details). The 5$\sigma$ detection limits determined in this way for all FIS
bands are listed in table \ref{tab_detlim}. These estimates also assume a
blackbody source with a temperature of 40\,K. We find that the estimates using
the simple method described in subsubsection \ref{sec:analy_est} and here agree
very well each other. The largest discrepancy occurs for the N60 band, where
the estimated detection limit using scanning simulation is lower by around
10\,\%. The instrumental noise in ISO observation is estimated to be 15--45 mJy
(\cite{herb98}; \cite{dole01}). Assuming our background brightness of $\sim$ 5
MJy~sr$^{-1}$, this noise level is similar with our estimation in the wide
bands. We analyse the photometric accuracy of point sources in more realistic
simulations with distributed sources below.

    \begin{table}
    \begin{center}
    \caption{$5\sigma$ detection limits of FIS bands from
      scanning simulations with a single pixel.}
       \vskip 3 truemm
    \begin{tabular}{l c } \hline\hline\noalign{\smallskip}
     & Detection Limits (mJy) \\
    \hline\noalign{\smallskip}
     WIDE-L& 40  \\
     N170 & 80 \\
     WIDE-S & 20  \\
     N60 & 47  \\
    \hline
    \end{tabular} \label{tab_detlim}
    \end{center}
    \end{table}

\subsection{Simulations with Distributed Point Sources and Realistic
Detector Configurations} \label{section-confusion} The FISVI takes into account
the full configuration of FIS detector arrays. We now discuss the simulations
over a finite patch of the sky with randomly distributed sources. By carrying
out the photometry of simulated images, we should be able to determine more
realistic detection limits.

Most faint sources to be observed by the ASTRO-F/FIS are expected to be distant
galaxies. Since the size of the PSFs at far-infrared wavelengths is relatively
large, we expect that the number of sources overlapped within a given PSF will
be larger. In such a situation, the source confusion would be important for
faint sources. In this section, we consider how the source confusion would
affect the observations by the ASTRO-F/FIS.


\subsubsection{Source distribution}\label{sec:sour_dist}
The effect of confusion depends on the distribution of sources in the sky and
the PSF. We assume that $N(>S)$, the number of sources whose flux is greater
than flux $S$, as a power-law on $S$,
\begin{equation}
N(>S) = N_0 (> S_0) \left({S\over S_0}\right)^{-\gamma},
\label{eq_sdist}
\end{equation}
for $S_{\rm min} < S < S_{\rm max}$, where $N_0$ and $S_0$ are normalisation
constants. For uniformly distributed sources in Euclidean space, $\gamma$ is
$1.5$. If the galaxies experience strong luminosity evolution from active to
less active star formation with time, $\gamma$ will become greater than 1.5.
The curved space could also give $\gamma$ different from 1.5. The analysis of
IR galaxy counts by ISO and SCUBA suggests that $\gamma$ would be greater than
1.5 but lower than 2.5 at around $\sim$ 150 mJy (\cite{puget99}; \cite{fran01};
\cite{pear01}; \cite{dole01}). Matsuhara et al. (2000) suggested that $\gamma$
could be steeper than 2.5 based on the fluctuation analysis due to the strong
evolution. In this paper, we examine three cases: $\gamma=1.5$, 2.5, and 3.0.
We fixed $S_{\rm min}$ = 10 mJy throughout the paper. Since there is no
divergence due to $S_{\rm max}$, we do not fix this number.

We need to specify the normalisation constants, $N_0$, at a given flux $S_0$,
which is set to be 100 mJy. These constants are determined from IR galaxy
counts normalised to Euclidean law [$N(>S) \propto S^{-1.5}$] at 90 $\mu$m
based on the IRAS survey and the European Large Area ISO Survey (\cite{efst00};
\cite{fran01}). In the following cases, though the source count results are
different for different bands and galaxy evolution, we assumed that there are
10 sources brighter than 100 mJy per square degree, i.e., $N_0$($>$ 100 mJy) =
10, in every observational band and the SED of all sources are flat within a
given FIS band. The number density of sources was estimated to be 316 per
square degrees corresponding to 0.2 within a circle of radius of $W_{\rm H}$ in
LW bands for $\gamma=1.5$ with the above normalisation. The density becomes
10-times larger for the case of $\gamma=2.5$ and the case of $\gamma=1.5$ and
$N_0$ = 100, and 19 times larger for the case of $\gamma=3.0$ and $N_0$ = 60.
We expect that source confusion becomes important for these distribution. The
distribution of sources in the sky is assumed to be uniform Poisson. In this
work, we want to check the pure confusion effect for the same distributed
galaxies by excluding other factors, e.g., various types of SED, the redshift
distribution, the luminosity function and the galaxy evolution. For a
comparison, we also check other cases: the Euclidean space with a large
normalisation constant ($N_0$ = 100) and an extreme case ( $\gamma=3.0$, $N_0$
= 60) (\cite{matsuh00}).

\subsubsection{Simple estimate of the confusion noise}\label{sec:sim_conf}
Although the clustering of sources could also affect the confusion noise, we
ignore such a possibility for simplicity. Following Condon (1974) and
Franceschini (1989), we obtain the noise due to confusion as

\begin{equation}
\sigma_{\mathrm{confusion}}^2 = \int_0^{x_{\mathrm{c}}} x^2 R(x) dx,
\label{eq-conf_sigma}
\end{equation}
where $x$ [$=S\,h(\theta,\phi)$] is the intensity, $x_{\mathrm{c}}$ is a cutoff
value, and $R(x)$ is the mean number of sources within the normalised beam
pattern, $h(\theta,\phi)$:
\begin{equation}
R(x) = \int_{\Omega_{\rm beam}} n\left({x\over
h(\theta,\phi)}\right)
 {d\Omega\over h(\theta,\phi)},
\end{equation}
where $n(S)$ is a differential number count.

\begin{figure}
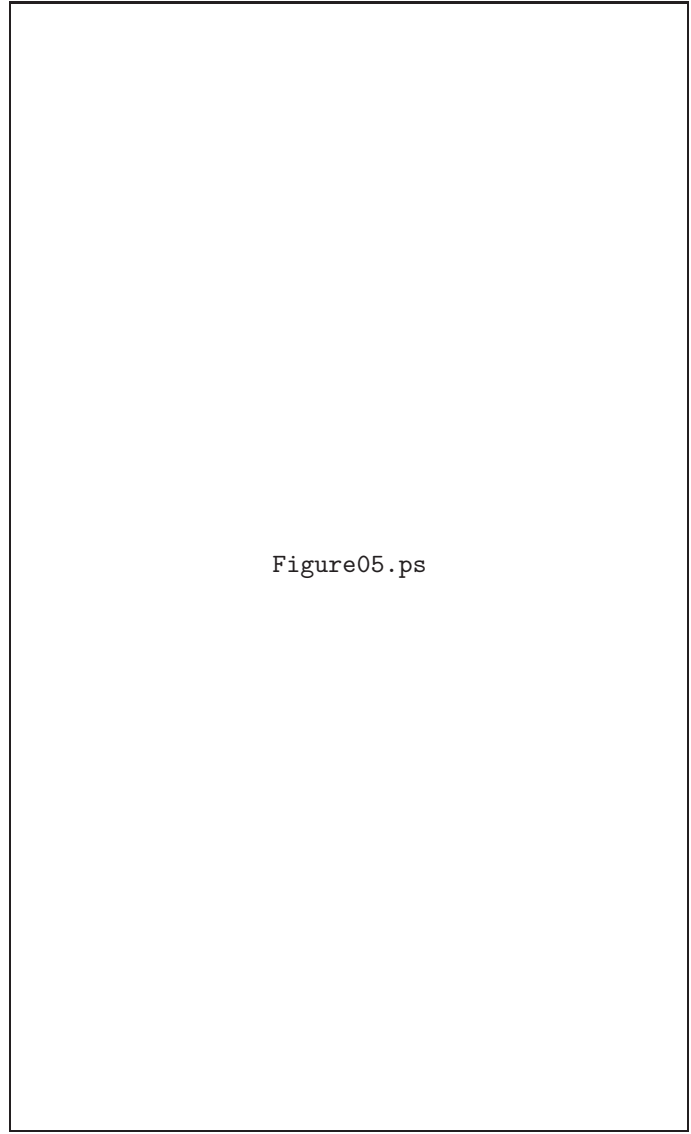

  \begin{center}
    \FigureFile(90mm, 150mm){Figure05.ps}
  \end{center}
     \caption{Comparison between the Compiled PSF (dotted) and the beam
     pattern (solid line) used in calculating the
     theoretical confusion (WIDE-L). Because we assumed a flat SED
     for all sources in this simulation, we used one Compiled PSF in
     the PSF-convolution.}
     \label{fig_drf}
\end{figure}

In this calculation, we use the beam pattern (see figure
\ref{fig_drf}), which is obtained from a simulated image of an
isolated point source using the FISVI without noises. The beam
pattern obtained in this way is somewhat wider than the Compiled
PSF due to pixel convolution. We also use the differential number
count obtained from the same source distribution assumed in
subsubsection \ref{sec:sour_dist}. These considerations are for
the purpose of comparing with the results from the photometry in
subsection \ref{sec:conf_phot}. We list the 5$\sigma$ confusion
noise in table \ref{tab_conf_analy}, obtained by using equation
(\ref{eq-conf_sigma}) for $\gamma=1.5$, $\gamma=2.5$, and
$\gamma=3.0$. We also estimated the crowded fields for
$\gamma=1.5$ by simply increasing $N_0$ by a large factor, i.e.,
$N_0(> 100$ mJy$) = 100$. The 5$\sigma$ confusion noise is the
same for the wide and narrow bands, because the beam patterns are
similar for two bands. Because of differences in the size of beam
profiles between long and short wavelengths, the detection limits
for LW are higher than those of SW bands. The detection limit by
confusion is approximately proportional to $N_0^{1/\gamma}$.

The confusion noise in FIRBACK survey by ISO is estimated to be around
$\sigma_{\rm c}$ $\simeq$ 45 mJy (\cite{dole01}). In our case, we used the
slope of the source distribution as $\gamma=1.5$ or $\gamma=2.5$ and set the
normalisation constant as $N_0$($>$ 100 mJy) = 10 by using the 90 $\mu$m source
count result (\cite{efst00}). Though the slope of the source count by Dole et
al. (2001) is similar to the Euclidean space ($\gamma=1.5$), the normalisation
constant should be different because the source density and the galaxy
evolution is different in other bands. Therefore, these discrepancies result
from the different normalisation and the cutoff flux ($S_{\rm min}$ = 10 mJy).

    \begin{table}
    \begin{center}
    \caption{$5\sigma$ detection limits due to confusion noise based on
     theoretical estimates.}
       \vskip 3 truemm
    \begin{tabular}{l c c c c} \hline\hline\noalign{\smallskip}
     & $\gamma=1.5$ & $\gamma=1.5$ & $\gamma=2.5$ & $\gamma=3.0$ \\
     & $N_0^{\mathrm{*}}$ = 10 & $N_0^{\mathrm{*}}$ = 100
     & $N_0^{\mathrm{*}}$ = 10 & $N_0^{\mathrm{*}}$ = 60 \\
     Band & (mJy) & (mJy) & (mJy) & (mJy) \\
    \hline\noalign{\smallskip}
     WIDE-L& 23 & 108 & 50 & 196 \\
     N170 & 24 & 115 & 52 & 204 \\
     WIDE-S & 12 & 54 & 35 & 123 \\
     N60 & 11 & 52 & 34 & 121 \\
    \hline
    \end{tabular} \label{tab_conf_analy}
    \begin{list}{}{}
    \item[$^{\mathrm{*}}$] $N_0(> 100 \rm\,mJy)$. Number per square degree.
    \end{list}
    \end{center}
    \end{table}

\subsection{Realistic Simulations}
\label{sec:conf_phot} The assumed source distribution of equation
(\ref{eq_sdist}) can be used to simulate the observed sky by the ASTRO-F/FIS.
By analysing the simulated images, we can address the effects of the various
sources of noises to the observation in a more realistic way.

\subsubsection{Realistic simulations}
\begin{figure}
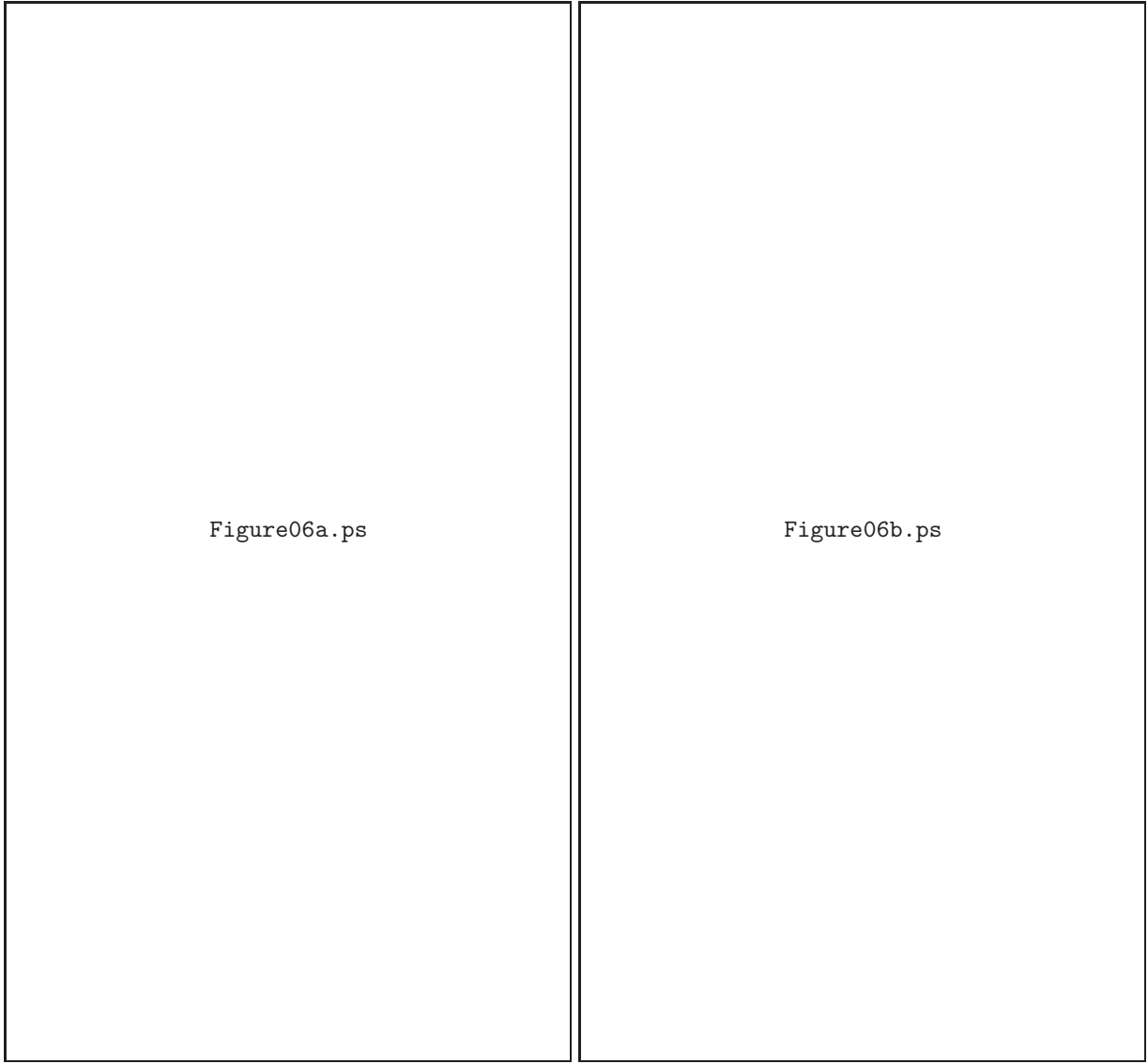

  \begin{center}
    \FigureFile(80mm, 150mm){Figure06a.ps}
    \FigureFile(80mm, 150mm){Figure06b.ps}
  \end{center}
   \caption{Simulated images in the WIDE-L band for distributed sources.
   We generated the distributed sources according to the cases of
   $\gamma=1.5$ and $N_0 = 10$ (left panel) and $\gamma=2.5$ and $N_0 = 10$
   (right panel). Photon and readout noises are added in these images.}
   \label{fig_sim_img}
\end{figure}

Using the FISVI, we generated two-dimensional images in the FIS
bands for two different virtual sky data with different $\gamma$.
We made two different sets of simulations. One was with the noise
levels described in earlier in this section; the other was with
the noise reduced to almost a negligible level in order to
separate the effects of confusion. The image size for the
distributed source simulation is $8192\arcsec \times 8192\arcsec$.
As mentioned in the previous section, we expect that the confusion
is important, especially for the cases that $\gamma$ is greater
than 2.5. In figure \ref{fig_sim_img}, we show an example of the
simulated images with the normal level of noise.

We carried out aperture photometry on the simulated images using SExtractor
software \textit{v}2.0.0 (\cite{bert96}). Some influential parameters were
optimised for better detection of the source, while the remaining were left
intact as default values. We set the threshold in the source detection and the
analysis as 3, the size of the photometric aperture as FWHM of beam pattern,
and we did not apply a filter for detection. In order to calibrate the output
flux, we used the five brightest input sources.

Figure \ref{phot_fr_none} shows the distribution of the $S_{\rm
out}/S_{\rm in}$ as a function of $S_{\rm in}$, where $S_{\rm in}$
and $S_{\rm out}$ denote the input flux and the flux obtained by
photometry. In the upper-left panel of figure \ref{phot_fr_none},
we assumed that it is for the case with $\gamma=1.5$ and $N_0=10$,
and negligible contribution of photon and readout noise. We also
assumed that detected source corresponds to the input source if
the position of the detected source lies within 9\arcsec\ for SW
bands and 15\arcsec\ for LW bands from the input source location.
We found very good correlation between the input and output
fluxes, and hence can conclude that the confusion noise is also
negligible for this case.

The noise added results for the case with $\gamma=1.5$ are shown in the
upper-left panel of figure \ref{phot_fr_noise}. The flux uncertainty becomes
significant near the estimated detection limits due to photon and readout
noise. Below the detection limit, most of the detected sources have an output
flux greater than the input flux: This is simply because detection can be
possible only when positive noises have been added to the source.

\begin{figure}
  \begin{center}
    \FigureFile(80mm, 150mm){Figure07a.ps}
    \FigureFile(80mm, 150mm){Figure07b.ps}
    \FigureFile(80mm, 150mm){Figure07c.ps}
    \FigureFile(80mm, 150mm){Figure07d.ps}
  \end{center}
   \caption{Flux ratio between the input and the output fluxes for extracted
   and identified sources in LW bands for the case of $\gamma=1.5$ and $N_0=10$
   (upper left), the case of $\gamma=1.5$ and $N_0=100$ (upper right),
   the case of $\gamma=2.5$ and $N_0=10$ (lower left), and the case of
   $\gamma=3.0$ and $N_0=60$ (lower right) without photon and readout noise.
   The flux in the vertical lines is 5$\sigma$ confusion noises calculated
   from equation (\ref{eq-conf_sigma}).
   The dotted line is for the N170 band and the dashed line is for the WIDE-L band.
   $S_{\rm in}$ and $S_{\rm out}$ mean the input flux and the
   output flux, respectively. As the source confusion is severer, the flux is
   boosted even in the high flux value.}
   \label{phot_fr_none}
\end{figure}

\begin{figure}
  \begin{center}
    \FigureFile(80mm, 150mm){Figure08a.ps}
    \FigureFile(80mm, 150mm){Figure08b.ps}
    \FigureFile(80mm, 150mm){Figure08c.ps}
    \FigureFile(80mm, 150mm){Figure08d.ps}
  \end{center}
   \caption{Flux ratio between the input and output fluxes for extracted
   and identified sources in LW bands for the case of $\gamma=1.5$ and $N_0=10$
   (upper left), the case of $\gamma=1.5$ and $N_0=100$ (upper right),
   the case of $\gamma=2.5$ and $N_0=10$ (lower left), and the case of
   $\gamma=3.0$ and $N_0=60$ (lower right) with the photon and readout noise.
    See the caption to figure \ref{phot_fr_none} for the meanings of
   the lines and symbols. In the case of including the photon and readout noise,
   the flux ratio is scattered near the detection limits by photon and readout
   noise. However, the trend of the boosted flux is similar to the case without noises
   (see figure \ref{phot_fr_none}), due to the heavy confusion.}
   \label{phot_fr_noise}
\end{figure}

The results with more crowded sources (i.e., $\gamma=2.5$ and $\gamma=3.0$) are
shown in the lower panels of figure \ref{phot_fr_none} for negligible noises,
and figure \ref{phot_fr_noise} for normal noises. Even with negligible noises,
we find that there are large deviations of the output fluxes from the input
fluxes. Thus, the flux uncertainties are mostly caused by the source confusion
shown in figure \ref{phot_fr_none}. Similar to the case dominated by the photon
and readout noise shown in the upper-right panel of figure \ref{phot_fr_noise},
$S_{\rm out}$ is systematically overestimated for sources below the theoretical
confusion limits. Such an upward bias was caused by source confusion; many of
the detected sources contain fainter sources within the beam. Actually, the
significant upward bias is partially due to the parameter, i.e., threshold, set
in SExtrator. First, SExtractor estimates the background fluctuation from each
local area. Because we reduce the noise below a negligible level, the
calculated background fluctuations are mainly due to many dim sources. The
detected sources at low flux surely have a flux above the fluctuation times the
threshold; these detected sources cause a significant upward bias. In the case
of heavy confusion, the trend of the boosted flux (see the lower panels of
figure \ref{phot_fr_noise}) is very similar to the case without noises (see the
lower panels of figure \ref{phot_fr_none}), which means that the faint sources
work as the dominant noise.

\begin{figure}
  \begin{center}
    \FigureFile(80mm, 150mm){Figure09a.ps}
    \FigureFile(80mm, 150mm){Figure09b.ps}
    \FigureFile(80mm, 150mm){Figure09c.ps}
    \FigureFile(80mm, 150mm){Figure09d.ps}
  \end{center}
   \caption{$N(>S)$ as a function of $S$ for the case of $\gamma=1.5$ and
   $N_0$ = 10 (upper left), the case of $\gamma=1.5$ and
   $N_0$ = 100 (upper right), the case of $\gamma=2.5$ and
   $N_0$ = 10 (lower left), and the case of $\gamma=3.0$ and
   $N_0$ = 60 (lower right) with the photon and readout noise.
   $N(>S)$ is the number of sources whose flux is greater than S in the size
   of the simulated image ($8192\arcsec \times 8192\arcsec$).
   The black solid lines represent the `true (or input)' distribution
   and symbols show the `observed' results. The vertical lines are the same
   as figure \ref{phot_fr_none}. The bend at low S is mainly due to the
   detection limit dominated by photon and readout noise. Also, the source
   confusion makes the slope significantly steeper than the true distribution
   in the case of the lower panels.}
   \label{fig_sc_noise}
\end{figure}

Figure \ref{fig_sc_noise} shows the integrated source count results. For a
comparison, we also plot the input source distribution. In the case of weak
source confusion (i.e., $\gamma=1.5$ and $N_0 = 10$) (upper-left panel of
figure \ref{fig_sc_noise}), the source count from a simulated image follows the
input source distribution well, except for the faint ends dominated by photon
and readout noise. However, the lower panels of figure \ref{fig_sc_noise} show
that the source distribution deviated from the input one due to source
confusion. The location of the estimated confusion limit of table
\ref{tab_conf_analy} is also shown in this figure. The observed slope is
significantly different from the input slope. The output slope can be 1.5-times
larger than the input slope in the case of a crowded source distribution.

As we mentioned in subsubsection \ref{sec:sim_conf}, we generated crowded
fields for the case of $\gamma=1.5$ by simply increasing $N_0$ by a large
factor, i.e., $N_0(> 100$ mJy$) = 100$ and the case of $\gamma=2.5$ and $N_0$ =
10, to exclude photon and readout noise in order to check the effect of pure
source confusion. Because there are no significant difference between the case
with and without the photon and readout noise, as can be seen in the
upper-right panel of figures \ref{phot_fr_none} and \ref{phot_fr_noise}, we
show the source count result with the photon and readout noise in the
upper-right panel of figure \ref{fig_sc_noise} in comparison with the
less-crowded case (upper left). Clearly, the confusion becomes important at
around $S = 100$ mJy for WIDE-L according to a theoretical calculation, but the
slope does not change. The change in the slope appears to occur only when the
underlying $N(>S)$ varies rather steeply on $S$. The lower-right panel of
figure \ref{fig_sc_noise} shows the case of $\gamma=3.0$ and $N_0$ = 60,
including the photon and readout noise. The slope of the source count is
significantly changed by the heavy confusion, and the source detection mainly
depends on source confusion.

\subsubsection{Detection limits from simulations}
It is not easy to define the detection limits from the simulated data. Since
the detection becomes increasingly difficult for sources below the detection
limits, we first define the `detection correctness' such that the ratio of the
number of correctly detected sources to the number of detected sources from the
photometry. We assume that the flux of the correctly detected source is the
measured flux from the photometry, and agrees with the input flux within a 20\%
error. The detection correctness can be near unity for sources well beyond the
detection limit, and goes down rapidly below the detection limit. We find that
the detection correctness reaches around 0.7 at the estimated detection limit
of a single scan. We thus define the location of the 70\% detection correctness
as the detection limit in our simulated data.

Figure \ref{fig_det_ratio} show a plot of the detection correctness with the
photon and readout noise. We first attempted to estimate the detection limit
purely due to source confusion. We arbitrarily suppressed the photon and
readout noise by a factor of 100 so that the noise-dominated detection limit
would become much less than the lower limit of the source flux of 10 mJy. The
resulting detection limits, estimated based on the detection correctness, are
summarized in table \ref{tab_conf_phot_none}. Under this condition, because the
source detection is affected by the source confusion and the photometric
accuracy, we could obtain similar detection limits in both narrow and wide
bands. These numbers are similar to those in table \ref{tab_conf_analy}, except
for $\gamma=1.5$, where the detection correctness remains larger than 0.7, even
for the faintest sources and for the case of the crowded source distribution.
This means that the confusion is not important for $\gamma=1.5$ and $N_0(> 100$
mJy$) = 10$.

    \begin{table}
    \begin{center}
    \caption{Detection limits for distributed point sources without
    photon and readout noise.}
       \vskip 3 truemm
    \begin{tabular}{l c c c c} \hline\hline\noalign{\smallskip}
     & $\gamma=1.5$ & $\gamma=1.5$ & $\gamma=2.5$ & $\gamma=3.0$ \\
     & $N_0^{\mathrm{*}}$ = 10 & $N_0^{\mathrm{*}}$ = 100
     & $N_0^{\mathrm{*}}$ = 10 & $N_0^{\mathrm{*}}$ = 60 \\
     Band & (mJy) & (mJy) & (mJy) & (mJy) \\
    \hline\noalign{\smallskip}
     WIDE-L & no confusion & 100 & 58 & 355 \\
     N170 & no confusion & 105 & 61 & 390 \\
     WIDE-S & no confusion & 45 & 31 & 305 \\
     N60 & no confusion & 40 & 30 & 278 \\
    \hline
    \end{tabular} \label{tab_conf_phot_none}
    \begin{list}{}{}
    \item[$^{\mathrm{*}}$] $N_0(> 100 \rm\,mJy)$. Number per square degree.
    \end{list}
    \end{center}
    \end{table}

\begin{figure}
  \begin{center}
    \FigureFile(80mm, 150mm){Figure10a.ps}
    \FigureFile(80mm, 150mm){Figure10b.ps}
    \FigureFile(80mm, 150mm){Figure10c.ps}
    \FigureFile(80mm, 150mm){Figure10d.ps}
  \end{center}
   \caption{Detection correctness for distributed sources with photon
   and readout noise for the case of $\gamma=1.5$ and $N_0$ = 10 (top left),
   the case of $\gamma=1.5$ and $N_0$ = 100 (top right), the case of $\gamma=2.5$
   and $N_0$ = 10 (bottom left), and the case of $\gamma=3.0$ and $N_0$ = 60
   (bottom right). The detected ratio is the ratio of the number of correctly
   detected sources (within a 20\% error) to the number of detected sources
   from the photometry. A detected ratio of 1.0 means all detected sources
   have been correctly detected. The detected ratio for the case of
   weak confusion (i.e., $\gamma=1.5$ and $N_0$ = 10) rapidly approaches 1.0
   in all bands. However, due to heavy confusion, the detected ratio does not
   approach 1.0 in the case of $\gamma=3.0$ and $N_0$ = 60.}
   \label{fig_det_ratio}
\end{figure}

Table \ref{tab_conf_phot_noise} shows the estimates of combined detection
limits where the readout noise, the photon noise, and the confusion noise are
considered. Since the confusion is not important for the case of $\gamma=1.5$
and $N_0$ = 10, the detection limit is purely determined by the photon and
readout noise. For the case of $\gamma=1.5$ and $N_0$ = 100, $\gamma=2.5$, and
$\gamma=3.0$, both the source confusion and the other noises contribute to the
detection limits. The combined detection limits for this case exceeds both the
noise dominated result (table \ref{tab_analy_est}) and source confusion
dominated result (table \ref{tab_conf_analy}). In the case of $\gamma=3.0$ and
$N_0$ = 60, we cannot exactly determine the detection limits because the severe
confusion makes the source detection difficult. Too many sources (i.e.,
$\gamma=1.5$ and $N_0$ = 100, $\gamma=3.0$ and $N_0$ = 60) also act as the
large amount of the photon noise, which affects in raising the detection limit.
Therefore, accurate photometry could be an additional important factor for
approaching the theoretical confusion limit in these cases.

    \begin{table}
    \begin{center}
    \caption{Detection limits for distributed point sources with photon
    and readout noise, taking account of the effects of the performance of
    the entire system, the brightness of the sky, the telescope
    emission, and the distribution of sources.}
       \vskip 3 truemm
    \begin{tabular}{l c c c c} \hline\hline\noalign{\smallskip}
     & $\gamma=1.5$ & $\gamma=1.5$ & $\gamma=2.5$ & $\gamma=3.0$ \\
     & $N_0^{\mathrm{*}}$ = 10 & $N_0^{\mathrm{*}}$ = 100
     & $N_0^{\mathrm{*}}$ = 10 & $N_0^{\mathrm{*}}$ = 60 \\
     Band & (mJy) & (mJy) & (mJy) & (mJy) \\
    \hline\noalign{\smallskip}
     WIDE-L & 26 & 125 & 68 & 440 \\
     N170 & 66 & 135 & 115 & 442 \\
     WIDE-S & 21 & 82 & 40 & 310 \\
     N60 & 49 & 92 & 63 & 280 \\
    \hline
    \end{tabular} \label{tab_conf_phot_noise}
    \begin{list}{}{}
    \item[$^{\mathrm{*}}$] $N_0(> 100 \rm\,mJy)$. Number per square degree.
    \end{list}
    \end{center}
    \end{table}

Matsuhara et al. (2000) analysed the ISO data obtained for the high density
case ($\gamma=3.0$ and $N_0$ = 60) from the fluctuation analysis method, which
is different from our photometric method. Because they assumed that the
fluctuation is mainly caused by unresolved faint point sources, they could
count the number of sources, even in a low flux range.

\section{Summary}
We have written observing simulation software, `FISVI', for an upcoming
infrared survey mission, ASTRO-F. Utilizing this software, we have estimated
the performance of the Far Infrared Surveyor (FIS) onboard ASTRO-F for ideal
conditions. We can carry out scanning simulations with a reasonable amount of
computing resources by introducing the Compiled PSF. The software can be used
to generate virtual data sets for a data-reduction pipeline.

We estimated the detection limits under various circumstances. For the case of
a non-crowded source distribution, the readout noise is usually more important
than the photon noise for dark patches of the sky by a factor of 1.3 to 2.5.
This means that the bright parts of the sky can be easily dominated by photon
noise. The emission from the telescope is less than the interstellar background
as long as the telescope temperature remains less than 6 K, but it could
contribute significantly to the long-wavelength band if the temperature becomes
larger than 6.5 K (see figure \ref{bgr_kawada}).

In crowded fields, source confusion becomes important in identifying sources.
The detection correctness becomes smaller for fainter sources. We have defined
the confusion limit in such a way that the number of correctly detected sources
within a 20\% error becomes larger than 70\% of the number of detected sources
from photometry. Such a definition of the confusion-dominated detection limit
gives very similar values of the confusion limit based on a simple formula. The
source confusion becomes larger than the detection limits by photon and readout
noise only if the number of faint sources becomes much larger than a simple
extension of the IRAS source counts down to around 10 mJy, assuming no
luminosity or density evolution. Recent models of source counts based on ISO
and SCUBA observations (\cite{matsuh00}; \cite{dole01}; \cite{fran01};
\cite{pear01}), however, predict the source distribution that is subject to
significant confusion at the longest wavelength band (WIDE-L). Other bands
appear to be noise-limited. The source confusion also could change the slope in
$\log N$--$\log S$ plots.

In this paper, we have made many simplifying assumptions concerning the sky
conditions. The actual sky brightness varies from place to place. The overall
statistics of the galaxy counts should be significantly influenced by
irregularities of the sky backgrounds. Also, in order to understand
cosmological effects, we will consider various types of SED, the luminosity
function, and the redshift distribution. The current version of FISVI does not
take into account more complicated behaviors of the detectors. These issues
will be discussed in forthcoming papers.

\section*{Acknowledgment}
W.-S. Jeong, J. Sohn, and I. Ahn were financially supported by the
BK21 Project of the Korean Government. They also appreciate
hospitality while staying at ISAS. This work was financially
supported in part by the KOSEF-JSPS corporative program. We thank
Myungshin Im, Chris Pearson, and Glenn J. White for reading our
manuscript and giving many suggestions.

\appendix
\section{Compiled PSF}\label{sec:comp_PSF}

\subsection{PSF Convolution}

\begin{figure}
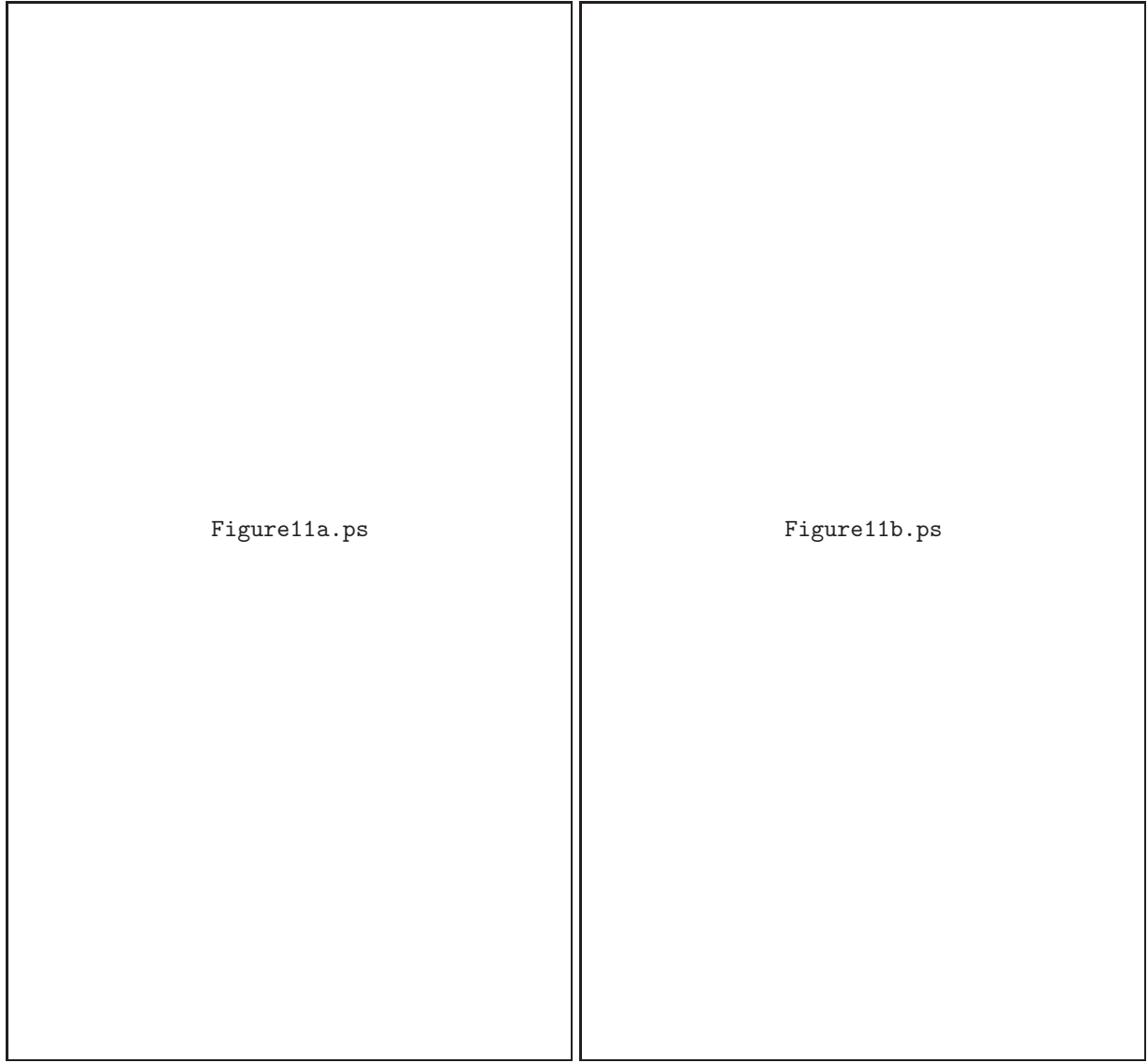

  \begin{center}
    \FigureFile(80mm, 150mm){Figure11a.ps}
    \FigureFile(80mm, 150mm){Figure11b.ps}
  \end{center}
  \caption{PSF of the ASTRO-F/FIS at 200 $\mu$m
     in a linear scale (left panel) and a logarithmic scale (right panel).
     The solid line shows the PSF simulated by using ZEMAX and the
     dotted line shows the Airy pattern with the assumption of a single
     circular aperture system.}\label{fig_psf}
\end{figure}

The PSF of ASTRO-F/FIS, including the entire optical path, was computed using
the ZEMAX optical simulation software package (Focus Software, Inc.). The
resulting PSF at $\lambda =200~\mu$m is shown in figure \ref{fig_psf}, together
with a circular aperture Airy pattern. The difference between the simulated PSF
and the Airy pattern is very small, but noticeable. The simulated PSF is
slightly narrower than the Airy pattern, and the side-lobe is more significant.
Since FIS detectors do not lie on the optical axis of ASTRO-F, the PSF is
slightly elongated with an ellipticity of $\sim 0.05$, but we assume the
circular PSF in the present simulations. Since the FIS covers a wide range of
wavelengths, the PSFs have been computed from 40 to 200 $\mu$m at 5 $\mu$m
intervals.

Using the simulated PSF, we first obtain the PSF-convolved image
$I_{\lambda,i}$ on the focal plane at wavelength $\lambda$, contributed solely
by the $i$-th point source:
\begin{equation}
I_{\lambda,i} (\textit{\textbf{r}})= F_{\lambda,i} \,
h_{\lambda}(\textit{\textbf{r}}; {\textit{\textbf{r}}_i^\prime})
\label{eq-single}
\end{equation}
and
\begin{equation}
1 = \int_{\Omega} h_{\lambda}(\textit{\textbf{r}};
{\textit{\textbf{r}}_i^\prime}) ~d\Omega, \label{eq-psf}
\end{equation}
where $\textit{\textbf{r}}$ is the position vector on the focal plane,
$F_{\lambda,i}$ is the flux density (at the wavelength $\lambda$) of the $i$-th
source, and $h_{\lambda}(\textit{\textbf{r}}; {\textit{\textbf{r}}_i^\prime})$
is the simulated PSF at wavelength $\lambda$ located centered at the position
of the $i$-th source $\textit{\textbf{r}}_i^\prime$. The PSF is normalised in
such a way that the integration over the entire solid angle becomes unity. The
intensity distribution on the focal plane, $I_{\lambda}(\textit{\textbf{r}})$,
can then be obtained by
\begin{equation}
I_\lambda (\textit{\textbf{r}} ) = \sum_i F_{\lambda,i} \, h_\lambda
(\textit{\textbf{r}}; \textit{\textbf{r}}_i^\prime). \label{eq_sums}
\end{equation}

\subsection{Filter Transmittance and Detector Response} As the
detector sweeps the sky, it integrates the charge generated by photons that
fall onto the detector. For a given intensity distribution on the focal plane,
$I_{\lambda} (\textit{\textbf{r}})$, the power,
$P_\lambda(\textit{\textbf{r}})$, at the wavelength interval $d\lambda$ is

\begin{equation}
P_\lambda(\textit{\textbf{r}})d\lambda = \int_{\Omega_{\rm pixel}}
I_{\lambda}(\textit{\textbf{r}}) \, A_{\rm tel} \, \tau(\lambda )~d\Omega
d\lambda, \label{eq_power}
\end{equation}
where $A_{\rm tel}$ is the effective collecting area of the
telescope, and $\tau(\lambda)$ is the filter transmittance along
the photon path within FIS (\cite{taka00}). The integration is
performed over the solid angle subtended by the pixel.

The detector transforms the photons into charges. The total
charge, $D$, integrated from $t_1$ to $t_2$ is

\begin{equation}
D(t_1 \rightarrow t_2) =\int_\lambda \int_{t_1}^{t_2} P_\lambda
(\textit{\textbf{r}}(t))\, \xi(\lambda) ~dt d\lambda, \label{eq_totcharge}
\end{equation}
where $\xi(\lambda)$ is the detector response function in units of A\,W$^{-1}$.
We use the following convention:
\begin{equation}
\xi(\lambda)\equiv \xi_0 \, \tilde \xi (\lambda),
\end{equation}
where $\xi_0$ is a constant in units of A\,W$^{-1}$ and $\tilde
\xi$ is a function normalised to unity at the peak value for SW
(short wavelength) and LW (long wavelength) detectors.

\begin{figure}
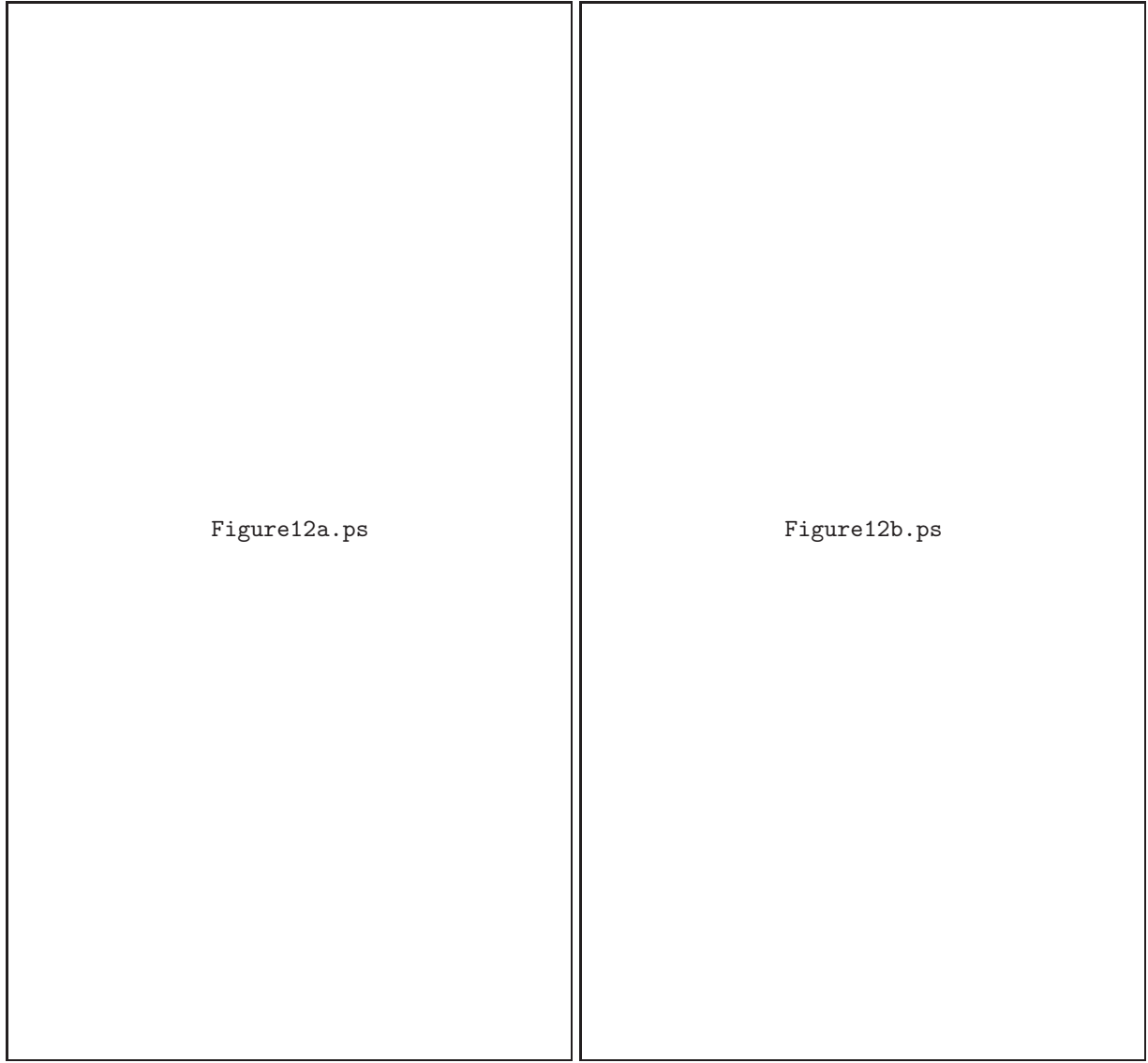

  \begin{center}
    \FigureFile(80mm, 150mm){Figure12a.ps}
    \FigureFile(80mm, 150mm){Figure12b.ps}
  \end{center}
    \caption{Filter transmission, $\tau(\lambda)$, (dashed lines) and the detector's
    response function, $\tilde \xi(\lambda)$, (dotted lines) for the N60 band (left)
    and the Wide-S band (right). The combined responsivities are shown
    as solid lines in arbitrary units.} \label{res_sw}
\end{figure}

\begin{figure}
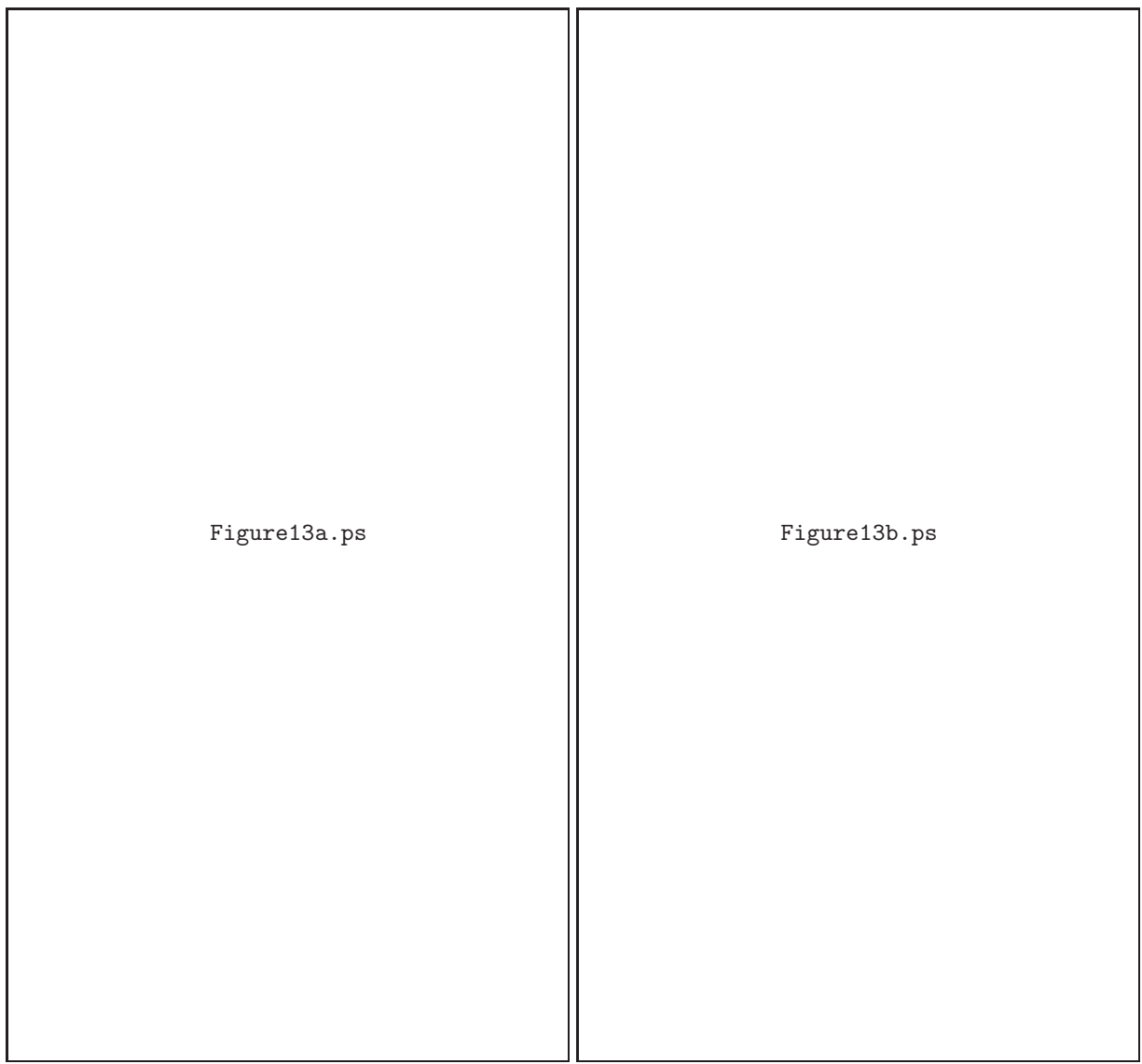

  \begin{center}
    \FigureFile(80mm, 150mm){Figure13a.ps}
    \FigureFile(80mm, 150mm){Figure13b.ps}
  \end{center}
    \caption{Same as figure \ref{res_sw}, except for
    the N170 band (left) and the WIDE-L band (right).}
    \label{res_lw}
\end{figure}

The normalised detector response functions, $\tilde \xi$, of LW and SW bands
are shown in figures \ref{res_sw} and \ref{res_lw}, respectively. We use these
curves and the measured detector responsivity, $\xi_{\rm{r}}$, to determine the
normalisation constant, $\xi_0$.
Measurements are done using a blackbody source, a filter that cuts off the
photons below a certain wavelength, a Winston cone, and a detector in a
perfectly reflecting cavity. The LW detector has long wavelength cut-off at 200
$\mu$m and SW detector at 110 $\mu$m. A low-pass filter was used to cut off the
photons at wavelength below the FIS band. The short wavelength limits were 140
$\mu$m for the LW detector and 40 $\mu$m for the SW detector. The measured
responsivity is represented by

\begin{equation}
{\xi_r} = \xi_0 \, {\int_{\lambda}\tilde \xi (\lambda) \,
B_\lambda(T) ~d\lambda \over \int_{\lambda} B_\lambda (T)
~d\lambda }, \label{eq_ms_resp}
\end{equation}
where $B_\lambda (T)$ is the Planck function at the temperature $T$. In this
estimation, we use $T=40$~K. From the measured value of $\xi_{\rm{r}}$ $\approx
20$~A\,W$^{-1}$ for LW, and $\xi_{\rm{r}}$ $\approx 7$ A\,W$^{-1}$ for SW, we
can determine the normalisation constant, $\xi_0$. The normalisation constants
are $\xi_0= 30$ A\,W$^{-1}$ for the LW and $\xi_0 =10$ A\,W$^{-1}$ for the SW
detectors, respectively.

\subsection{Compiled PSF}
If we use the same spectral energy distribution (SED) for each
source, the flux density of the source can be defined as
\begin{equation}
F_{\lambda, i} = \mathcal{F}_i \, S_\lambda, \label{eq_sed}
\end{equation}
where $S_\lambda$ is the spectral energy distribution (SED) normalised to unity
over the wavelength band and $\mathcal{F}_i$ is the flux integrated over the
bandwidth. We can rewrite equation (\ref{eq_sums}) as
\begin{equation}
I_\lambda (\textit{\textbf{r}} ) = \sum_i \mathcal{F}_{i} \, S_\lambda \,
h_\lambda (\textit{\textbf{r}}; \textit{\textbf{r}}_i^\prime). \label{eq_sums2}
\end{equation}
Since $\lambda$ is independent of $\textit{\textbf{r}}$ and
${\textit{\textbf{r}}_i^\prime}$, we can introduce a new function,
$H(\textit{\textbf{r}}; \textit{\textbf{r}}_i^\prime)$, by integrating over the
wavelength as
\begin{equation} H(\textit{\textbf{r}}; \textit{\textbf{r}}_i^\prime) = A_{\rm tel} \,
\int_\lambda h_\lambda (\textit{\textbf{r}}; \textit{\textbf{r}}_i^\prime) \,
S_\lambda \, \tau(\lambda) \, \xi(\lambda) ~d\lambda . \label{eq_cpsf}
\end{equation}
We define this $H(\textit{\textbf{r}}; \textit{\textbf{r}}_i^\prime)$ as the
`Compiled PSF'. If we perform convolution to the image plane by using this
Compiled PSF, we can avoid repeated wavelength integration. Finally, equation
(\ref{eq_totcharge}) can be rewritten as
\begin{equation}
D(t_1 \rightarrow t_2) = \int_{t_1}^{t_2}\int_{\Omega_{\rm pixel}} \sum_i
\mathcal{F}_{i} H(\textit{\textbf{r}}; \textit{\textbf{r}}_i^\prime) ~ d\Omega
dt. \label{eq_totcpsf}
\end{equation}
This concept of the Compiled PSF is effective only when the number of SED type
is limited. The calculation time is reduced by a factor of $N_\lambda$ by using
the Compiled PSF, where $N_\lambda$ is the number of wavelength grids. With a
wavelength interval of $\Delta\lambda = 5~\mu$m, a typical $N_\lambda$ lies
between 10 and 20. In order to carry out simulations over four square degrees
in the WIDE-S band, we need about 15 hours of computing time with Pentium IV 1
GHz machines. By introducing Compiled PSF, we can accomplish such a simulation
within an hour.

\subsection{Spectral Energy Distribution of the Sources}

We expect that the majority of faint point sources detected by the ASTRO-F/FIS
will be external galaxies. Each object will have its own SED, but most
extragalactic point sources in the infrared band can be classified into four
types of galaxies, i.e., the cirrus type representing typical spiral galaxies,
the M 82 type starbursts, the Arp 220 type starbursts and the AGN dust torus
type (Rowan-Robinson 2001). Four Compiled PSFs are required to accommodate
these four types of SEDs in the simulations. The observed SEDs are further
affected by the redshifts. We need redshifted-dependent SEDs for each type of
source.

\begin{figure}
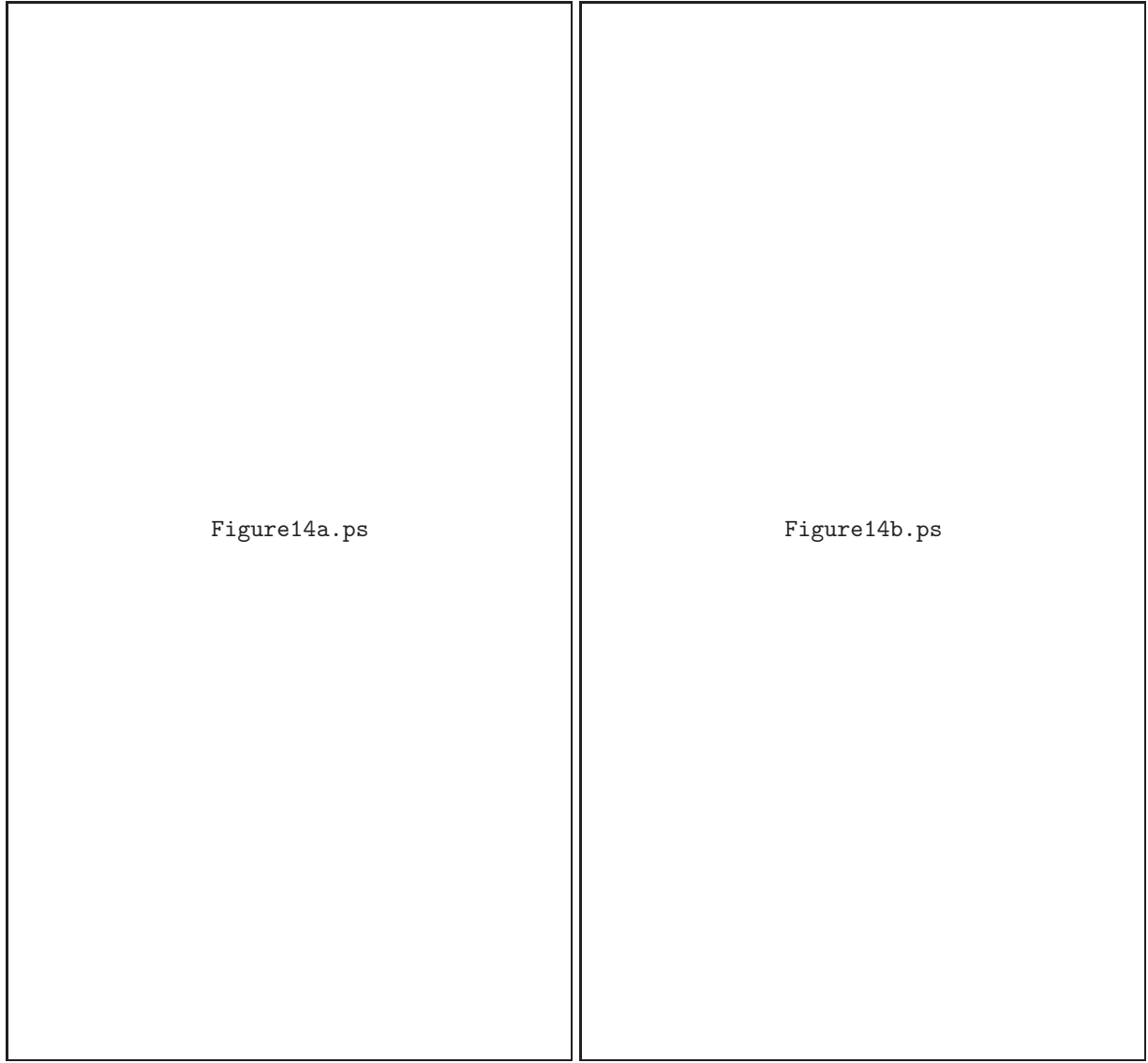

  \begin{center}
    \FigureFile(80mm, 150mm){Figure14a.ps}
    \FigureFile(80mm, 150mm){Figure14b.ps}
  \end{center}
    \caption{Normalised Compiled PSFs in the WIDE-L band.
    The left panel is the Compiled PSFs over SED for redshift 1.0 and
    the right panel is the Compiled PSFs over redshift for the cirrus
    type. For a comparison, we also plot the Compiled PSF computed
    for galaxies with the flat SED used in this work.}
    \label{fig_seds}
\end{figure}

We expect that the Compiled PSF will be changed with the SED types and the
redshift for wide bands, but the difference was found to be very small, even
for the WIDE-S and WIDE-L bands, as shown in figure \ref{fig_seds}. Since our
main purpose is to examine the general performance of the ASTRO-F/FIS, we
concentrate on simple models for the nature of the sources. We will deal with
the SED types of sources, redshift distributions, and the luminosity function
in the next paper in order to understand the cosmological model and the galaxy
evolution through the observing simulation. Though the difference between the
Compiled PSFs computed from the flat SED and other SEDs is severe at some
extreme cases ($\sim$ 10\% difference over the area), we use the Compiled PSF
computed for galaxies with the flat SED in the present paper (i.e., $F_\lambda$
= constant) (see figure \ref{fig_seds}). In the flat SED's case, the Compiled
PSF does not depend on the redshift.

\section{Procedures of Scanning and Data Sampling}\label{sec:pos}

\begin{figure}
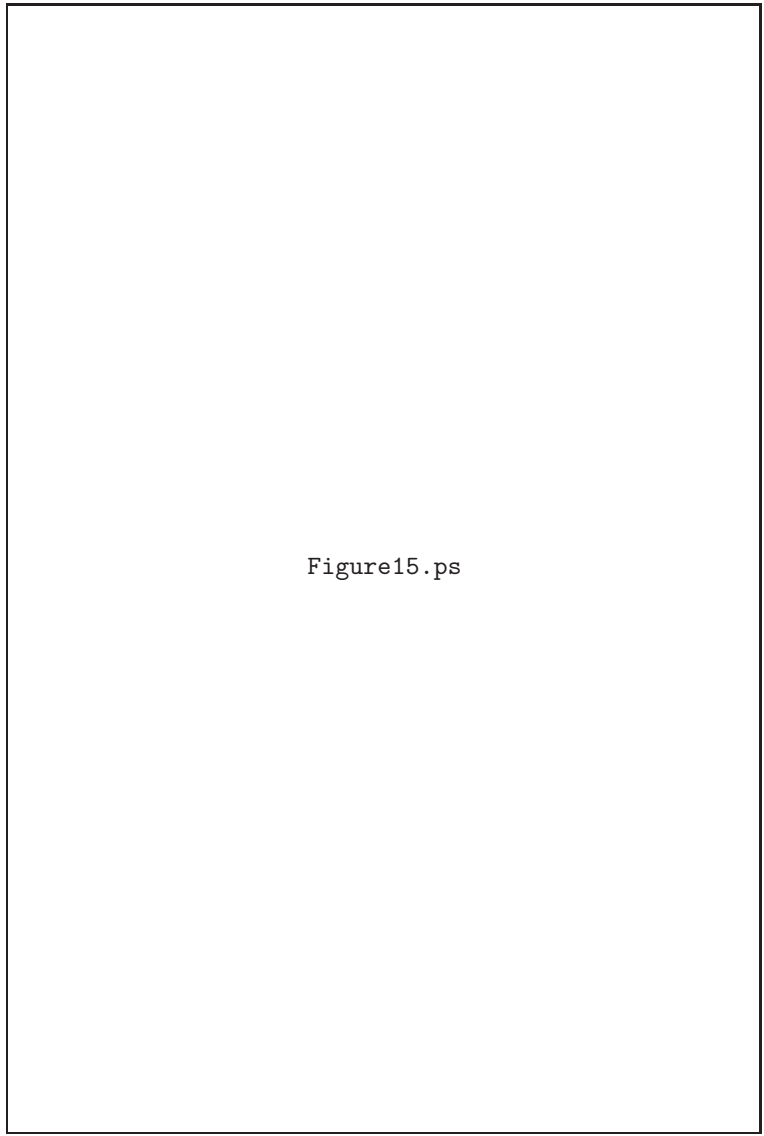

  \begin{center}
    \FigureFile(100mm, 150mm){Figure15.ps}
  \end{center}
   \caption{Layout of the detector array for the N170 band and definition of
    the scan directions.}
   \label{fig_scan_dir}
\end{figure}

\begin{figure}
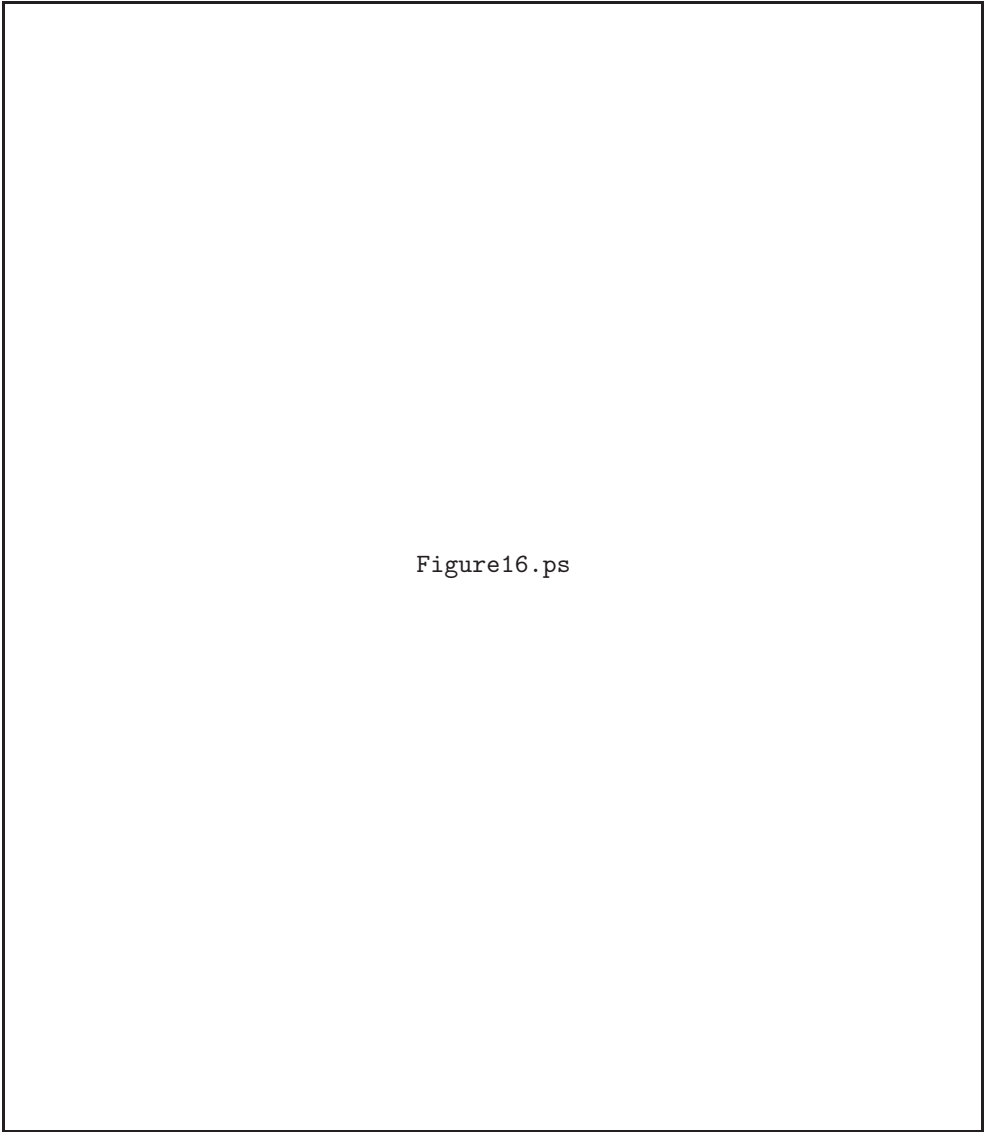

  \begin{center}
    \FigureFile(130mm, 150mm){Figure16.ps}
  \end{center}
   \caption{Passage of the detector for the WIDE-L band. We
   display the footprints of the detector pixels scanned three times.}
   \label{fig_pos_det}
\end{figure}

A PSF-convolved image is generated on grids where the scanning procedure is
performed. To scan a PSF-convolved image, we need to know the position of the
detector pixels. We set the array of the starting point to scan on the $x$
(cross-scan direction) and the $y$ (in-scan direction) frame in the image. The
FIS detector arrays have 2 or 3 rows and 15 or 20 columns, and is tilted by an
angle $\theta = 26.^\circ$5 from the cross-scan direction in order to assure
Nyquist sampling (\cite{taka00}; \cite{matsu01}). We denote $i$ as the index
for the sampling sequence, and $j$ and $k$ as the indices for the row and
column of the detector array, respectively (see figure \ref{fig_scan_dir}). By
denoting ($x_0$, $y_0$) as the position vector of the center of upper left
pixel of the array at the beginning of the scan (i.e., $i=j=k=0$), we have the
following formulae for the position vectors of the $(j,k)$ pixel at the
$(i+1)$-th sampling:
\begin{equation}
x(i,j,k) = x_0 + p \left(k \, \cos\theta  + j \, \sin\theta\right)
\end{equation} {\rm and }
\begin{equation}
y(i,j,k) = y_0 + i\,v\Delta t + p \left( j \, \cos\theta - k \,
\sin\theta\right) ,
\end{equation}
where $p$ is the size of the pixel pitch (see table \ref{tab_spec_fis}), $v$ is
the scanning angular speed (which is 3.60 arcmin~s$^{-1}$) of the satellite,
and $\Delta t$ is the increment of the detector motion in the scan direction
during the sampling interval. Note that the $x$ position of each pixel does not
depend on $i$ in this coordinate system. We show one example for the passage of
the detector in figure \ref{fig_pos_det}.

\section{Integrating over the Detector Pixel}
\label{sec:sampling}

The integration of equation (\ref{eq_totcpsf}) over $\Omega_{\mathrm{pixel}}$
was carried out by summing up the image convolved with Compiled PSF on fine
grids. The image convolved with Compiled PSF was constructed on grids of $4''$
resolution, but the accuracy of the $\Omega_\mathrm{pixel}$ integration was not
good enough on such grids ($\sim$ a few percent error), partly because of the
tilted configuration of the detector arrays. In order to improve the accuracy
of the integration, we laid finer grids over the area where the integration
would be performed. We were able to reduce the integration error down to 1\% by
taking a three-times finer grid over the integration area. If we use a smaller
grid, we can improve the accuracy of the flux and the position further, but we
would need more computing time.

\begin{figure}
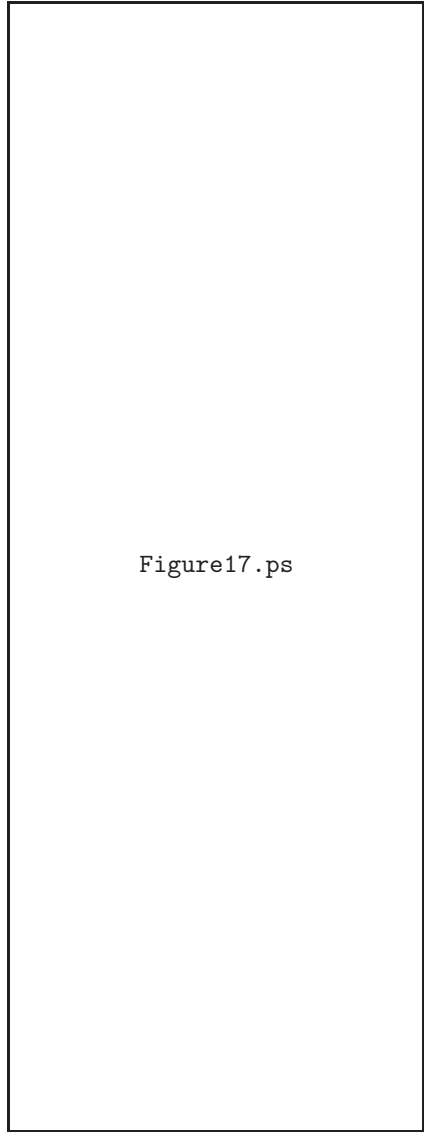

  \begin{center}
    \FigureFile(55mm, 150mm){Figure17.ps}
  \end{center}
   \caption{Each sampling is composed of several subsamples in
   order to ensure accurate integration over the region where
   the intensity varies. s1--s5 mean the subsamples. In actual
   simulations, we used only two subsamples.}
   \label{fig_int_pixel}
\end{figure}

The time integration of equation (\ref{eq_totcpsf}) was made by
dividing one sampling interval to shorter subsampling intervals in
order to mimic the continuous scanning of the detector and
applying the trapezoidal rule to the subsampled time series data.
As the detector moves, one detector pixel integrates the signal
during the subsample interval (see figure \ref{fig_int_pixel}).
The number of subsample determines the resolution of integrated
signal values. The sampling rate of 15.2 Hz for LW bands
corresponds to 14\farcs2 which is much smaller than the pixel
size, and we found that we need only two subsamples to ensure the
integration accuracy over time becomes smaller than 1\%.

\begin{thebibliography}{}

\bibitem[Bertin, Arnouts(1996)]{bert96}
Bertin, E., \& Arnouts, S.\ 1996, \aaps, 117, 393

\bibitem[Boggs, Jean(2001)]{boggs01}
Boggs, S. E., \& Jean, P.\ 2001, \aap, 376, 1126

\bibitem[Condon(1974)]{cond74}
Condon, J. J.\ 1974, \apj, 188, 279

\bibitem[Dole et al.(2001)]{dole01}
Dole, H., et al.\ 2001, \aap, 372, 364

\bibitem[Efstathiou et al.(2000)]{efst00}
Efstathiou, A., et al.\ 2000, \mnras, 319, 1169

\bibitem[Franceschini et al.(1989)]{fran89}
Franceschini, A., Toffolatti, L., Danese, L., \& De Zotti, G.\ 1989, \apj, 344,
35

\bibitem[Franceschini et al.(2001)]{fran01}
Franceschini, A., Aussel, H., Cesarsky, C. J., Elbaz, D., \& Fadda, D.\ 2001,
\aap, 378, 1

\bibitem[Garcia et al.(1998)]{garc98}
Garcia, R. A., Roca Cort$\acute{\rm e}$s, T., \& R$\acute{\rm e}$gulo, C.\
1998, \aaps, 128, 389

\bibitem[Herbstmeier et al.(1998)]{herb98}
Herbstmeier, U., et al.\ 1998, \aap, 332, 739

\bibitem[Jeong et al.(2000)]{jeong00}
Jeong, W.-S., et al.\ 2000, in Proc. of Mid- and Far-Infrared Astronomy and
Future Missions, ed. T. Matsumoto \& H. Shibai, ISAS Report, SP14, 297

\bibitem[Kawada(1998)]{kawada98}
Kawada, M.\ 1998, \procspie, 3354, 905

\bibitem[Kawada(2000)]{kawada00}
Kawada, M.\ 2000, in Proc. of Mid- and Far-Infrared Astronomy and Future
Missions, ed. T. Matsumoto \& H. Shibai, ISAS Report, SP14, 273

\bibitem[Kiss et al.(2001)]{kiss01}
Kiss, C., $\acute{\rm A}$brah$\acute{\rm a}$m, P., Klaas, U., Juvela, M., \&
Lemke, D.\ 2001, \aap, 379, 1161

\bibitem[Matsuhara et al.(2000)]{matsuh00}
Matsuhara, H., et al.\ 2000, \aap, 361, 407

\bibitem[Matsuura et al.(2001)]{matsu01}
Matsuura, M., Nakagawa, T., Murakami, H., \& Yamamura, I.\ 2001, ISAS Report,
681, 1

\bibitem[Murakami(1998)]{mura98}
Murakami, H.\ 1998, \procspie, 3356, 471

\bibitem[Nakagawa(2001)]{naka01}
Nakagawa, T.\ 2001, in the Proc of The Promise of the Herschel Space
Observatory, ed. G. L. Pilbratt, J. Cernicharo, A.M. Heras, T. Prusti, \& R.
Harris, ESA-SP, 460, 67


\bibitem[Pearson(2001)]{pear01}
Pearson, C. P.\ 2001, \mnras, 325, 1511

\bibitem[Puget et al.(1999)]{puget99}
Puget, J. J., et al.\ 1999, \aap, 345, 29

\bibitem[Rieke(1994)]{rieke94}
Rieke, G. H.\ 1994, in Detection of Light: from the Ultraviolet to the
Submillimeter, ed. K. Visnorsky (Cambridge: Cambridge University Press), pp.
65-67

\bibitem[Rowan-Robinson(2001)]{MRR01}
Rowan-Robinson, M.\ 2001, \apj, 549, 745

\bibitem[Shibai(2000)]{shib00}
Shibai, H.\ 2000, in IAU Symp. 204, The extragalactic background and its
cosmological implications, ed. M. Harwit \& M. G. Hauser (Michigan:
Astronomical Society of the Pacific), 455

\bibitem[Takahashi et al.(2000)]{taka00}
Takahashi, H., et al.\ 2000, \procspie, 4013, 47

\end{thebibliography}
\end{document}